 \documentclass[sigconf]{acmart}
\AtBeginDocument{%
  \providecommand\BibTeX{{%
    \normalfont B\kern-0.5em{\scshape i\kern-0.25em b}\kern-0.8em\TeX}}}

\setcopyright{acmcopyright}
\copyrightyear{2024}
\acmYear{2024}
\setcopyright{rightsretained}
\acmConference[CHI EA '24]{Extended Abstracts of the CHI Conference on Human Factors in Computing Systems}{May 11--16, 2024}{Honolulu, HI, USA}
\acmBooktitle{Extended Abstracts of the CHI Conference on Human Factors in Computing Systems (CHI EA '24), May 11--16, 2024, Honolulu, HI, USA}
\acmDOI{10.1145/3613905.3650823}
\acmISBN{979-8-4007-0331-7/24/05}
\usepackage{acmart-taps}
\title[How Does Conversation Length Impact User's Satisfaction]{How Does Conversation Length Impact User's Satisfaction? A Case Study of Length-Controlled Conversations with LLM-Powered Chatbots}

\usepackage{xspace}
\usepackage{comment}
\usepackage{booktabs}

\newcommand{\eg}{{\it e.g.}\xspace}

\usepackage{dirtytalk}
\usepackage{amsmath}
\usepackage{booktabs,siunitx}
\usepackage{enumitem}
\usepackage{caption}
\usepackage{subcaption}
\usepackage{cleveref}
\usepackage{xcolor}
\usepackage{multirow}
\usepackage{graphicx}

\newenvironment{choruschat}
    {
        \begin{enumerate}[leftmargin=3pc,style=nextline,align=right,itemsep=0mm]
        \small
        \sffamily
        \vspace{.3pc}
    }
    {
        \vspace{.3pc}
        \end{enumerate}
    }
\newcommand{\chatgptAssistant}[1]{\item [assistant:]\xspace #1} %assistant chat
\newcommand{\chatgptUser}[1]{\item [\textbf{user:}]\xspace \textbf{#1}} %study2
\newcommand{\chatgptSystem}[1]{\item [system:]\xspace #1}
\newcommand{\assistantA}[1]{\item [MultiSlack:]\xspace #1}
\newcommand{\assistantB}[1]{\item [SlackVanilla:]\xspace #1}

\newcommand{\slackBaseline}{\textsc{SlackVanilla}\xspace}
\newcommand{\slackBot}{\textsc{MultiSlack}\xspace}

\aptLtoX[graphic=no,type=html]{\newenvironment{myquote}{\begin{quote}}{\end{quote}}}{\newenvironment{myquote}%
  {\list{}{\leftmargin=0.1in\rightmargin=0.1in}\item[]}%
  {\endlist}}
\usepackage{cleveref}

\begin{document}

\author{Shih-Hong Huang}
\affiliation{%
  \institution{The Pennsylvania State University}
  \city{University Park}
  \state{PA}
  \country{USA}}
\email{szh277@psu.edu}

\author{Ya-Fang Lin}
\affiliation{%
  \institution{The Pennsylvania State University}
  \city{University Park}
  \state{PA}
  \country{USA}}
\email{yml5563@psu.edu}

\author{Zeyu He}
\affiliation{%
  \institution{The Pennsylvania State University}
  \city{University Park}
  \state{PA}
  \country{USA}}
\email{zmh5268@psu.edu}

\author{Chieh-Yang Huang}
\affiliation{%
  \institution{The Pennsylvania State University}
  \city{University Park}
  \state{PA}
  \country{USA}}
\email{chiehyang@alumni.psu.edu}

\author{Ting-Hao `Kenneth' Huang}
\affiliation{%
  \institution{The Pennsylvania State University}
  \city{University Park}
  \state{PA}
  \country{USA}}
\email{txh710@psu.edu}

%%
%% By default, the full list of authors will be used in the page
%% headers. Often, this list is too long, and will overlap
%% other information printed in the page headers. This command allows
%% the author to define a more concise list
%% of authors' names for this purpose.
\renewcommand{\shortauthors}{Huang et al.}

%%
%% The abstract is a short summary of the work to be presented in the
%% article.
\begin{abstract}
Users can discuss a wide range of topics with large language models (LLMs), but they do not always prefer solving problems or getting information through lengthy conversations.
This raises an intriguing HCI question: How does instructing LLMs to engage in longer or shorter conversations affect conversation quality?
In this paper, we developed two Slack chatbots using GPT-4 with the ability to vary conversation lengths and conducted a user study. 
Participants asked the chatbots both highly and less conversable questions, engaging in dialogues with 0, 3, 5, and 7 conversational turns. 
We found that the conversation quality does not differ drastically across different conditions, while participants had mixed reactions.
Our study demonstrates LLMs' ability to change conversation length and the potential benefits for users resulting from such changes, but we caution that changes in text form may not necessarily imply changes in quality or content.
\end{abstract}

%%
%% The code below is generated by the tool at http://dl.acm.org/ccs.cfm.
%% Please copy and paste the code instead of the example below.
%%


%%
%% Keywords. The author(s) should pick words that accurately describe
%% the work being presented. Separate the keywords with commas.

%% A "teaser" image appears between the author and affiliation
%% information and the body of the document, and typically spans the
%% page.
%\begin{teaserfigure}
%  \includegraphics[width=\textwidth]{sampleteaser}
%  \caption{Seattle Mariners at Spring Training, 2010.}
%  \Description{Enjoying the baseball game from the third-base
%  seats. Ichiro Suzuki preparing to bat.}
%  \label{fig:teaser}
%\end{teaserfigure}

%\received{20 February 2007}
%\received[revised]{12 March 2009}
%\received[accepted]{5 June 2009}

%%
%% This command processes the author and affiliation and title
%% information and builds the first part of the formatted document.
\maketitle

% \kenneth{You can use these custom commands to add comments}

\section{Introduction}
%\kenneth{------------------ KENNETH IS WORKING HERE: BEGINS ------------}

People can now talk about almost anything via conversation with large language models (LLMs).
However, it is known that users do not always want to solve their problems or obtain information through lengthy conversations.
A study surveying 500 AskReddit\footnote{https://www.reddit.com/r/AskReddit/} questions~\cite{huang2023whattype} showed that although people are enthusiastic about addressing specific types of questions through conversation, they prefer \textit{not} to tackle certain questions, like those about food or technology problems, through conversations.
Moreover, individuals often find themselves in situations that hinder them from engaging in extended conversations, such as walking or cooking. 
They may simply need a quick answer and move on.

This raises an intriguing HCI question: \textbf{How does instructing LLMs to engage in longer or shorter conversations affect conversation quality?}
Answering this question is challenging because a conversation necessitates human user involvement, and the unfolding of the conversation cannot be examined by solely testing the LLMs.
To keep this research question manageable, this paper focuses on two specific situations:
{\em (i)} What happens if LLMs were instructed to engage in longer conversations with a human user for less conversable questions?
Conversely, {\em (ii)} what if LLMs were instructed to hold a short conversation for questions that people prefer to discuss in longer conversations?

To answer this question, we developed two Slack\footnote{https://slack.com/} chatbots using GPT-4 with the ability to vary conversation lengths and conducted a user study. 
Participants were directed to ask both more and less conversable questions to the chatbots, engaging in dialogues with 0, 3, 5, and 7 conversational turns. 
We collected participants' self-reported responses and enlisted online crowd workers to rate the quality of these conversations.
The study results show that while GPT-4 can alter conversation lengths, this does not necessarily lead to improved conversation quality. Participants' self-reported satisfaction levels varied and are said to be dependent on different situations. According to online crowd workers' ratings, the conversation quality does not vary drastically across different conditions.
Having longer conversations for highly conversable questions did not result in better quality conversations, and having shorter conversations for less conversable questions did not result in worse quality conversations.
However, participants had mixed reactions: some found that longer conversations were more engaging and offered more nuanced discussions, while others considered longer conversations repetitive and not useful.

Our study demonstrated LLMs' ability to change conversation length, thus offering versatile applications for information.
Additionally, our study cautions that changes in text form may not necessarily imply changes in quality or content.

\section{Backgrounds}
% \input{Sections/2-related-work}

%\section{\system System}
%\input{Sections/3-scicapenter-system}

%Do users' circumstances influence their preferred input method with conversational assistants?
%Does adjusting conversation length based on the question enhance user satisfaction with conversational assistants?

% \section{Two Curious Case Studies of Situations and Conversational Assistants}

%\kenneth{Need to say something here about why we tested on these situations, as situations are hard to test if we don't have an application.}

% \section{Study 1 - Patrol Study: Do users' situations influence their preferred input method?}
% \label{sec:study-1}
% \kenneth{TODO Alan: Move the walking study to here.}

% Patrol Study: Effects of Using Voice Interfaces to Receive Remote Help
% \input{Sections/4-study1/4-study1}

% \section{Results}
% \input{Sections/4-study1/4-study1-results}

% Ken: 2. HCI papers (say, from CHI, CSCW, UIST, etc.) that studied users' reactions or perspectives for different lengths (or formats, or styles) of a piece of text or a conversation. Namely, they take a piece of text and modify its length, styles, or format *without changing its meaning too much* and test user's reaction to it. The texts do not necessarily need to be generated by machines; the texts can be human-written.

% Different form factors, modifications
% motivation: benefits of appropriate linguistic choices
Employing suitable linguistic styles may augment perceptions of trustworthiness~\cite{morrissey2013realness, tallyn2018ethnobot} and anthropomorphism~\cite{gnewuch2017towards, hill2015real}, as well as improve the perceived quality of the interactive experience \cite{jakic2017impact}. 
HCI researchers have extensively explored the impact of linguistic style choices on user perceptions of conversational agents. However, little research explores the influence of the length of conversational turns.
% persona 
% Among the studies for users' perceptions of linguistic style choices, a prevalent focus is user's perceptions of persona-based design approaches~\cite{hwang2019data, park2023audilens}. 
% In these approaches, the linguistic style of a chatbot is strategically crafted to align with the distinct traits of the target persona, as noted in studies~\cite{hwang2019data, park2023audilens}.

% influence of style
Several research endeavors have been undertaken to empirically assess the impact of various conversational styles on user experiences (UX) with chatbots~\cite{araujo2018living, chen2020creating, elsholz2019exploring, chaves2022chatbots}.
For example, \citet{chaves2022chatbots} conducted a study where parallel corpora were developed, consisting of dialogues within the tourism sector, which exhibited analogous content but differed in the situation of use. The research focused on assessing user preferences regarding the linguistic choices of chatbots and evaluating aspects such as appropriateness, credibility, and overall user experience. \citet{elsholz2019exploring} conducted a comparative analysis of user interactions with chatbots employing modern English versus those utilizing a Shakespearean style of language. Findings from this study indicated that the chatbot implementing modern English was regarded as more user-friendly, whereas the one using Shakespearean English was perceived as providing a more entertaining experience.
In research aimed at discerning the anticipatory needs of migrants in relation to information-seeking chatbots~\cite{chen2020creating}, it was observed by the participants that an overly informal and relaxed linguistic style was not congruent with the context of the subject matter.
Along the line of research, \citet{hu2018touch} designed a tone-aware chatbot that can generate responses in a passionate tone and empathetic tone. Users consider chatbot responses to be as passionate as human ones and even more empathetic than human ones.

% \subsubsection{Summary}
% [Q and A consent form] An AI-powered chatbot aiming to improve informed consent online by taking questions while going through the consent form. \cite{xiao2023inform}

However, this body of literature lacks the investigation of users' perception of chatbots' ability to navigate users' questions. In the real world, people explore complex, ill-defined questions together. Prior research has recognized the need for creating conversational systems that can help with ill-defined questions~\cite{huang2018evorus}. \citet{huang2023whattype} examined the types of ill-defined open-ended questions that are most suitable for having conversations. Furthermore, natural language processing (NLP) and AI researchers have studied the ability of large language models to control the length of the outputs~\cite{zhang2023survey}, such as headline length control~\cite{kanungo2022cobart}, the ability to extend conversations~\cite{sun2023plug}. We leverage the above research progress and aim to investigate whether chatbots' quantity of relevant conversational turns could support users' ill-defined questions.

\section{User Study}
\label{sec:study-2}
% \kenneth{In short, yes.}
%\alan{this is a note, need rephrase}
%Study 2, conversation length study aims to investigate whether the altering the conversation length according to the questions being asked affects the satisfaction level of users.
The study aimed to investigate the potential \textbf{user benefits of assistants that strategically adapt conversation formats}, especially regarding \textbf{conversation length}. 
ChatGPT enabled us to develop a chatbot capable of responding to arbitrary questions while varying the number of conversational turns it engages in.
We used questions compiled by~\citet{huang2023whattype} and further categorized them into high Conversation score and low Conversation score brackets. 
We hypothesized that users would benefit from longer conversations when posing highly-conversable questions.

\paragraph{Study Design.}
The main component of the study was a \textbf{conversational question-answering} session.
Participants were required to interact with two automated Slackbots\footnote{https://slack.com/resources/using-slack/a-guide-to-slackbot-custom-responses} of different configurations, \slackBot and \slackBaseline,
on Slack backed by ChatGPT (GPT-4) to help answer a set of questions.
\slackBaseline gives answers immediately, while \slackBot holds the conversation for several turns for clarification.
% \alan{justification for slack, different use case as whatsapp, sit in front computer}
% \alan{We chose Slack for its popularity, ease of integration with various external services, and the ability to export conversation logs for analysis.?}
Participants participated in person in the research lab using the provided computer and monitor.
%The author's institute's IRB office approved the study. 

%\alan{why choose slack: }

%\paragraph{Selecting the Questions and Preparing the Answers}
Questions used in the study were drawn from the \texttt{AskReddit Questions Dataset}~\cite{ask-reddit} annotated by~\citet{huang2023whattype}.
The questions were rated on various aspects, including their conversational nature (Conversation score) and the degree to which they prompt individuals to seek out others to pose the questions (Reach-Out score), among other factors.
We chose a total of 40 questions: 20 with the highest Conversation score ($\geq$ 4.11)
and 20 with the lowest Conversation score ($\leq$ 2.67).
% \avon{as Huang et. al. suggested?}.
In cases where the questions had the same Conversation score,
we prioritized those with a higher Reach-Out score. We also corrected minor typos in the selected questions. 

For the remainder of this paper, we will use the terms \textit{Conversability/Conversable} to refer to the Conversation score. Example questions include \textit{``Is it possible to make a good situation out of any bad situation?''} (high-conversability), and \textit{``Do you know any website to watch Marvel's ``What If...?''?''} (low-conversability).
% \avon{what is Q1?} 
We provide the 40 selected questions in \Cref{sec:question-pool}, where
% Lists of the 40 selected questions are shown in \Cref{tab:question-pool-high} and \Cref{tab:question-pool-low}.
%
\Cref{tab:question-pool-high} displays the 20 selected questions with highest Conversation scores, and \Cref{tab:question-pool-low} presents the 20 selected questions with lowest Conversation scores.

We excluded two categories of questions from our list.
The first group consisted of sensitive topics and Reddit-specific questions,
including those related to Not Safe For Work (NSFW) content,
religion (especially those of a sensitive or discriminative nature),
topics involving minors, and Reddit technical inquiries.
The second group comprises those to which ChatGPT cannot directly respond to.
This includes requests for web links, suggestions to harm others,
and inquiries regarding serious mental health issues.
%\kenneth{Can we add one example questions for high and low conversable questions?}\steven{done}

\paragraph{Configuring the Slackbots.}
In this study, we built two GPT-4 Slackbots: \slackBaseline and \slackBot.
\slackBaseline is configured to provide direct responses without engaging in iterative conversations.
Conversely, \slackBot is designed to initiate a dialogue, asking up to $N$ clarifying questions before delivering a final answer, where $N$ is a variable that will be manually set by researchers in the study.
This interactive capability is realized by inserting an additional prompt
that tracks the conversation's progress and informs GPT-4 about the remaining number of follow-up inquiries after each user input.
Once the limit of follow-up questions is reached,
GPT-4 is prompted to stop further asking questions and produce the final response.
The implementation details of the two Slackbots are described in \Cref{sec:appendix-slackbot-implementation}.
We also included conversations of the same question between participants and \slackBaseline, as well as with \slackBot when $N=3$, $N=5$, and $N=7$ in \Cref{sec:example-conversation-2}.
% \kenneth{TODO: Can we show an example conversation of the same question with N=0, 3, 5, 7 in the appendix?}

% \paragraph{Implementation Details of the two SlackBots}
% % \label{sec:slackbot-implementation}
% \input{Sections/4-study2/4-study2-implementation-details}

\paragraph{Study Procedure.}
The study consisted of a pre-study session and two interaction sessions (Session \slackBot and Session \slackBaseline).

\begin{enumerate}
    \item 
    The \textbf{pre-study session} introduced the study and included a short tutorial.

    \item
    \textbf{Session \slackBot} and \textbf{Session \slackBaseline}.
    At each interactive session, participants were asked to perform the information-seeking and ask \slackBaseline and \slackBot on Slack to help answer the assigned question.
    Users filled out a set of questionnaire questions after each question was answered by the assistant.

\end{enumerate}
The questionnaire questions are presented in \Cref{sec:survey}, where \Cref{tab:study2-survey-A} and \Cref{tab:study2-survey-B} display the  questionnaire questions for \slackBaseline and \slackBot, respectively.

Each participant will ask three questions to \slackBot and three questions to \slackBaseline,
where the six questions are distinct from one another.
The order in which participants engaged with \slackBot or \slackBaseline was randomized.
Note that participants would consecutively ask three questions to \slackBot before switching to \slackBaseline, or vice versa.
\slackBot was configured in three different ways: to ask 3, 5, and 7 follow-up questions
before providing answers to the participants' questions, denoted as $N=3$, $N=5$, and $N=7$, respectively.
Eight different participants conversed with each of these three different \slackBot configurations.
Participants were not informed about the difference in assistant configuration before the study.
The three values of $N$ were decided through the four pilot study sessions. It was concluded that configuring $N \ge 10$ easily exceeds 1.5 hours of study time and caused fatigue for participants, potentially compromising the overall quality of the study.

For $N=5$, six questions were randomly selected from the pool of 40 questions (sampled from the AskReddit dataset~\cite{huang2023whattype, ask-reddit}).
To ensure a more comprehensive coverage of the questions,
for $N=3$ and $N=7$, three questions were randomly chosen from the questions with the highest Conversation scores ($\geq$ 4.11)
and three from the questions with the lowest Conversation scores ($\leq$ 2.67).
Additionally, the order of the questions was randomized.

Each participant took 1 to 1.5 hours to complete the entire study and the participants were compensated with \$20.00.
This study was approved by the IRB office of the authors' institute.

% 6 questions for each participant, 8 people for N=3,5,7, three questions for both assistants, 3+3, randomly decide which first 
% N=3, randomly select 6 from 40, N=5,7 3 high Q6 score, 3 low Q6 score for better question pool coverage.

\paragraph{Participants.}
Participants were recruited through university mailing lists and personal networks.
% (participants were ID coded P1-P24 in the following paper).
A total of 24 individuals were recruited (participants were ID coded U1-U24 in this paper): fourteen males, nine females, and one prefer not to say.
Twenty-two of the participants were aged between 21 and 35; two participants were over 36 years old.
The majority of the participants were undergraduate and graduate students at the university.
Among all the participants, only one did not have prior experience using virtual assistants (\eg, Siri, Google Assistant, or Alexa).
All participants have prior experience interacting with ChatGPT.
Participants were informed that they would be interacting with a ChatGPT-backed assistant during the study.

\section{Findings}
% Please add the following required packages to your document preamble:
% \usepackage{booktabs}
% \usepackage{multirow}
\begin{table*}[]
\centering \small\def\arraystretch{1.15}
\begin{tabular}{@{}lllccccc@{}}
\toprule
\textbf{} & \textbf{} & \textbf{} & \textbf{\slackBaseline} & \textbf{} & \multicolumn{3}{c}{\textbf{\slackBot}} \\ \cmidrule{4-4} \cmidrule{6-8}
\textbf{} & \textbf{} & \textbf{Converasbility} & \textbf{No Conversation} & \textbf{} & \textbf{N=3} & \textbf{N=5} & \textbf{N=7} \\ \midrule
\textbf{} & \multirow{2}{*}{\textbf{\# Dialogue}} & \textbf{High} & 44 &  & 9 & 14 & 12 \\
\textbf{} &  & \textbf{Low} & 28 &  & 15 & 10 & 12 \\ \hline
\multirow{4}{*}{\textbf{Q$_A$}} & \multirow{2}{*}{\textbf{\begin{tabular}[c]{@{}l@{}}Should Ask More\\ (Strongly Disagree -\textgreater Strongly Agree)\end{tabular}}} & \textbf{High} & 3.55 &  & - & - & - \\
 &  & \textbf{Low} & 3.25 &  & - & - & - \\ \cmidrule{2-8}
 & \multirow{2}{*}{\textbf{\begin{tabular}[c]{@{}l@{}}Enough Count\\ (Not enough -\textgreater Too Many)\end{tabular}}} & \textbf{High} & - &  & 3.11 & 2.71$^\dagger$ & 3.75$^\dagger$ \\
 &  & \textbf{Low} & - &  & 3.33 & 3.60 & 3.42 \\ \hline
\multirow{2}{*}{\textbf{Q$_B$}} & \multirow{2}{*}{\textbf{\begin{tabular}[c]{@{}l@{}}Helpfulness\\ (Strongly Disagree -\textgreater Strongly Agree)\end{tabular}}} & \textbf{High} & 3.57 &  & 3.44 & 3.86 & 4.25 \\
 &  & \textbf{Low} & 4.39 &  & 4.07 & 4.20 & 4.17\\ \hline
\multirow{2}{*}{\textbf{Q$_C$}} & \multirow{2}{*}{\textbf{\begin{tabular}[c]{@{}l@{}}Satisfaction\\ (Very Dissatisfied -\textgreater Very Satisfied)\end{tabular}}} & \textbf{High} & 3.64$^\bigstar$ &  & 3.44 & 3.57 & 4.33$^\bigstar$ \\
 &  & \textbf{Low} & 4.36 &  & 4.13 & 4.20 & 4.17 \\ \bottomrule
\end{tabular}
\caption{Average ratings of the participant self-reported response. Rating pairs that passed the T-Test are denoted by $\dagger$ (p-value = 0.027) and $\bigstar$ (p-value = 0.041). We observed that increasing conversation turns would improve user satisfaction level for high-conversation-score questions.}
\label{tab:slackbot-result}
\end{table*}
% \input{tables/slackbot-turk}
% \steven{added t-test result}

\begin{figure*}
\centering
\raisebox{8.5mm}{\includegraphics[width=0.49\linewidth]{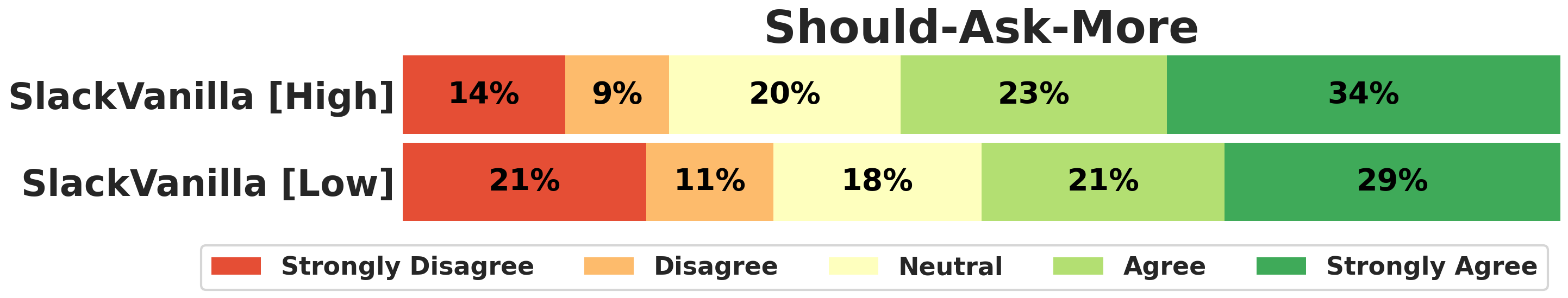}}%
\hfill%
\includegraphics[width=0.49\linewidth]{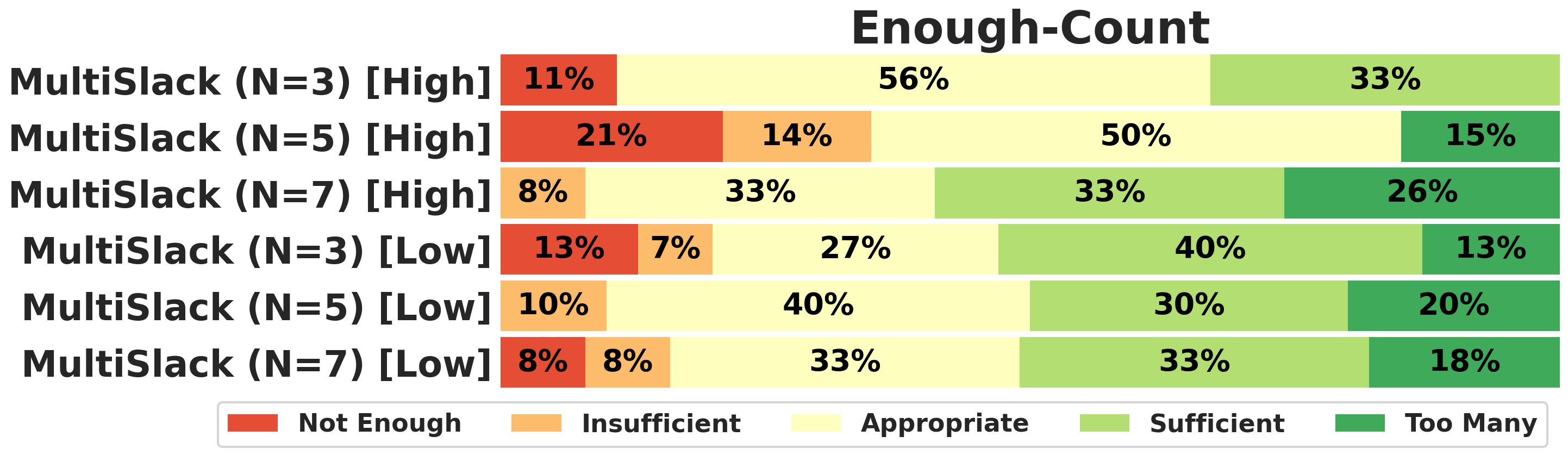}\\%
\includegraphics[width=0.49\linewidth]{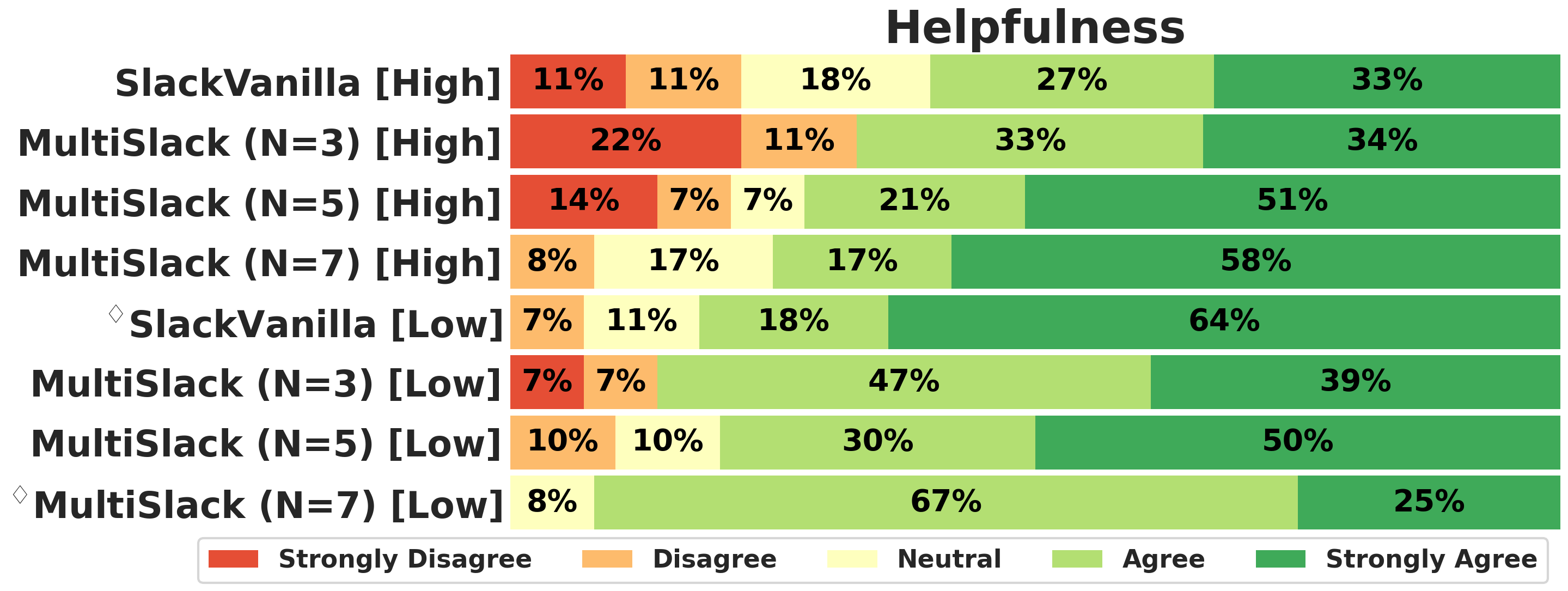}\hfill%
\includegraphics[width=0.49\linewidth]{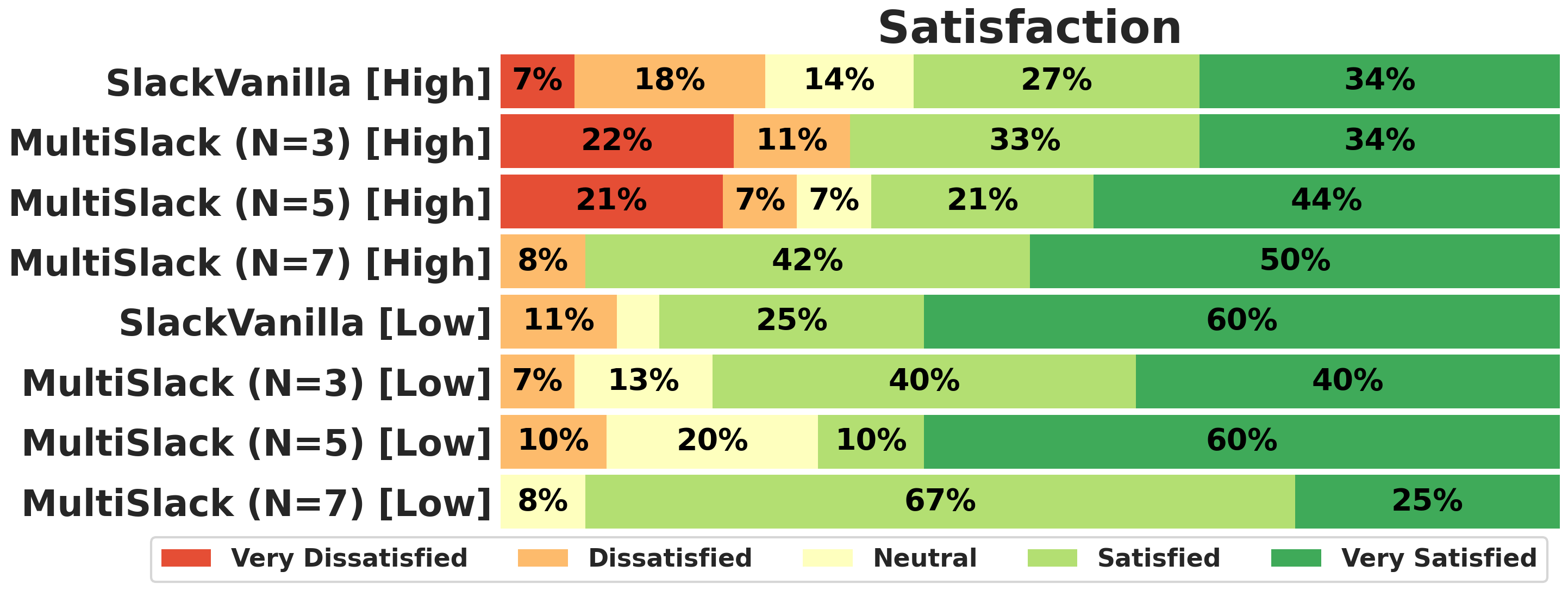}%
\caption{Rating distribution of the participant self-reported responses. Rating pair that passed the Chi-squared Test is denoted by $\diamondsuit$ (p-value = 0.05).}
    \Description{The figure contains four subfigures representing the distribution of the participant self-reported responses. Each subfigure is consisted of bar charts representing SlackVanilla and MultiSlack(N=3, 5, 7) for both high and low conversability questions. The bar charts are shown in five different colors to represent the five-point Likert scale, the colors used are red, orange, yellow, light green, and dark green.
    Should-Ask-More, located top left with red representing Strongly disagree, orange representing Disagree, yellow representing Neutral, light green representing Agree, and dark green representing Strongly agree. The two bar charts are 
    SlackVanilla [High]: 14\% red, 9\% orange, 20\% yellow, 23\% light green, 34\% dark green.
    SlackVanilla [Low]: 21\% red, 11\% orange, 18\% yellow, 21\% light green, 29\% dark green.
    Enough-Count, located top right with red representing Not Enough, orange representing Insufficient, yellow representing Appropriate, light green representing Sufficient, and dark green representing Too Much. The six bar charts are 
    MultiSlack (N=3) [High]: 11\% red, 56\% yellow, 33\% light green.
    MultiSlack (N=5) [High]: 21\% red, 14\% orange, 50\% yellow, 15\% dark green.
    MultiSlack (N=7) [High]: 8\% orange, 33\% yellow, 33\% light green, 26\% dark green.
    MultiSlack (N=3) [Low]: 13\% red, 7\% orange, 27\% yellow, 40\% light green, 13\% dark green.
    MultiSlack (N=5) [Low]: 10\% orange, 40\% yellow, 30\% light green, 20\% dark green.
    MultiSlack (N=7) [Low]: 8\% red, 8\% orange, 33\% yellow, 33\% light green, 18\% dark green.
    Helpfulness, located bottom left with red representing Strongly disagree, orange representing Disagree, yellow representing Neutral, light green representing Agree, and dark green representing Strongly agree. The eight bar charts are 
    SlackVanilla [High]: 11\% red, 11\% orange, 18\% yellow, 27\% light green, 33\% dark green.
    MultiSlack (N=3) [High]: 22\% red, 11\% orange, 33\% light green, 34\% dark green.
    MultiSlack (N=5) [High]: 14\% red, 7\% orange, 7\% yellow, 21\% light green, 51\% dark green.
    MultiSlack (N=7) [High]: 8\% orange, 17\% yellow, 17\% light green, 58\% dark green.
    diamondsuit, SlackVanilla [Low]: 7\% orange, 11\% yellow, 18\% light green, 64\% dark green.
    MultiSlack (N=3) [Low]: 7\% red, 7\% orange, 47\% light green, 39\% dark green.
    MultiSlack (N=5) [Low]: 10\% orange, 10\% yellow, 30\% light green, 50\% dark green.
    diamondsuit, MultiSlack (N=7) [Low]: 8\% yellow, 67\% light green, 25\% dark green.
    Satisfaction, located bottom right with red representing Very Dissatisfied, orange representing Dissatisfied, yellow representing Neutral, light green representing Satisfied, and dark green representing Very Satisfied. The eight bar charts are 
    SlackVanilla [High]: 7\% red, 18\% orange, 14\% yellow, 27\% light green, 34\% dark green.
    MultiSlack (N=3) [High]: 22\% red, 11\% orange, 33\% light green, 34\% dark green.
    MultiSlack (N=5) [High]: 21\% red, 7\% orange, 7\% yellow, 21\% light green, 44\% dark green.
    MultiSlack (N=7) [High]: 8\% orange, 42\% light green, 50\% dark green.
    SlackVanilla [Low]: 11\% orange, 4\% yellow, 25\% light green, 60\% dark green.
    MultiSlack (N=3) [Low]: 7\% orange, 13\% yellow, 40\% light green, 40\% dark green.
    MultiSlack (N=5) [Low]: 10\% orange, 20\% yellow, 10\% light green, 60\% dark green.
    MultiSlack (N=7) [Low]: 8\% yellow, 67\% light green, 25\% dark green.}
    \label{fig:distribution-self-reported}
\end{figure*}

% \alan{do we want to use high/low conversability or high/low conversation-score?}
% \cy{conversability sounds better for me!}
% \Cref{tab:slackbot-result}
\subsection{Participant Self-Reported Response Analysis}
\Cref{tab:slackbot-result} shows the average participant self-reported ratings for the questionnaire questions. \Cref{fig:distribution-self-reported} shows the rating distribution for the questionnaire questions. 
\subsubsection{As the conversation length increased, satisfaction levels for high-conversability questions also rose.}
In the post-interaction questionnaire, we asked participants to rate how satisfied they were with how the assistant helped them (Q$_C$) on a five-point Likert scale from Very Dissatisfied (1) to Very Satisfied (5).
As shown in \Cref{tab:slackbot-result} [Q$_C$, High], the average satisfaction scores for high-conversability questions were 3.64 for \slackBaseline, 3.44 for $N=3$, 3.57 for $N=5$, and 4.33 for $N=7$.
The average score of \slackBaseline and \slackBot $N=7$ show statistical difference (p-value = 0.041).
We observed an increase in the satisfaction score when the number of questions asked for the high-conversability questions increased.
These results affirmed our hypothesis: \textbf{assistants that strategically adjust conversation formats-- specifically conversation length-- in accordance with the current situation (the conversability of the question), offer advantages to users.}
% ([Q6, High, \slackBot with N=7] = 4.33 and [Q6, High, \slackBaseline] = 3.64).
% T test 
% Q6\_7A Q6\_7B pass t test 4.33 3.64 

% Q6(Satisfaction): Overall, are you satisfied with how the assistant help you solve the question? (Very dissatisfied >>> very satisfied)

\subsubsection{The helpfulness of responses to high-conversability questions increased with increasing conversation length.}
%\subsubsection{Helpfulness of Responses for High Conversation Score Questions Increases as Conversation Length Increases}
We also asked the participants how helpful the solution was (Q$_B$) on a five-point Likert scale from Strongly Disagree (1) to Strongly Agree (5).
As shown in \Cref{tab:slackbot-result} [Q$_B$, High],
the average score for high-conversability questions were 3.44 for $N=3$, 3.86 for $N=5$, and 4.25 for $N=7$.
Again, we observed an increase in the helpfulness score when increasing the number of questions asked for the high-conversbility questions.
These results, again, affirmed our hypothesis that \textbf{assistants that strategically tailor conversation formats to the current situation provide benefits to users.}

% Q5(Helpfulness): At the end of the conversation, the assistant provided a helpful solution to the question you initially asked. (Strongly Disagree >>> Strongly Agree)

%\subsubsection{Force Low Conversation Score Questions to Be Longer Effect Less Than Forcing High Conversation Score Questions to Be Shorter \alan{rephrase}}
% \subsubsection{Varying conversation length was less effective for low-conversability questions.}
% Extending the length of low-conversability questions had a smaller impact on the perceived quality across the board compared to shortening the high-conversability questions.
% We observe that low-conversability questions' score fluctuated less as the conversation length was changed (see the \textbf{Low} rows for \slackBot with $N=3$, $N=5$, and $N=7$ in \Cref{tab:slackbot-result}).
% High-conversability questions, on the other hand, were more sensitive to changes in conversation length (see the \textbf{High} rows for \slackBot with $N=3$, $N=5$, and $N=7$ in \Cref{tab:slackbot-result}).

% \alan{not sure yet}
\subsubsection{Participants may believe high-conversability questions necessitate more questions from the assistant.}
For \slackBaseline, we asked participants to rate whether the assistant should ask more questions (Q$_A$) on a five-point Likert scale from Strongly Disagree (1) to Strongly Agree (5).
The average score for the high-conversability questions is 3.55, and for the low-conversation-score questions is 3.25.

For \slackBot, we asked whether the number of questions asked by the assistant is enough (Q$_A$) on a five-point Likert scale from Not Enough (1) to Too many (5).
For the high-conversability questions, the average score of $N=7$ was 3.75, for $N=5$ is 2.71, and for $N=3$ is 3.1.
While average score of $N=7$ and $N=5$ show statistical differences (p-value = 0.027), the rise in N=3 score was not clear to the authors.

% Q4\_B: With the goal of answering your question sufficiently, do you think the assistant should ask more follow-up questions before providing its answer? (Strongly Disagree >>> Strongly Agree)
% Q4\_A: With the goal of answering your question sufficiently, how would you rate the number of follow-up questions the assistant asked before providing its answer? (Not enough >>> too many)

% Q4\_5A Q4\_7A pass t test 2.71 3.75
\subsubsection{\slackBot was preferred over \slackBaseline but with varying opinions from participants.}
We explicitly asked participants to compare \slackBot and \slackBaseline by asking them, \textit{``Regardless of the questions asked, how would you rate your interaction with the assistants?''} Participants used a five-point Likert scale ranging from \textit{``A Is Significantly Better''} (1) to \textit{``B Is Significantly Better''} (5). The average ratings and standard deviations for comparison between \slackBot of different configurations and \slackBaseline were as follows: for \slackBot with $N=3$ was (mean = 1.75, $\sigma$ = 0.97); for \slackBot with $N=5$ was (mean = 2.5, $\sigma$ = 1.5); and for \slackBot with $N=7$ was (mean = 2.25, $\sigma$ = 1.3). The high standard deviations suggest that participants held varying opinions. Additionally, many participants indicated that comparisons between \slackBot and \slackBaseline depended on specific factors. For example, 
U18 commented,
\begin{myquote}
\textit{``I don't think either of the bots particularly are better or worse than the other one so this is why I chose a 3. It felt like [SlackVanilla]'s direct responses to my questions were appropriate for the question types (typically ones that have factual/objective answers). For [MultiSlack], the questions were more focused on opinion/subjective topics, and I think its ability to provide follow-up questions is good for this case. I do think if you were to reverse the questions for the bots, the experience would be a lot different.''}
\end{myquote}
U12 said,
\begin{myquote}
\textit{``I cant choose which is better. It depends. Because [MultiSlack] may try to have a conversation with you while [SlackVanilla] will directly provide you the answers. I think sometimes I need to the answer quickly, I prefer [SlackVanilla]. Sometimes, my situation is complex and need to be analyzed carefully, I prefer [MultiSlack].''}
\end{myquote}
U14 reflected, 
\begin{myquote}
\textit{`` Time matters. if I'm in a rush to get a quick answer from a robot who does not have any follow-up question or empathy/emotion, and then I would prefer [SlackVanilla]. However, if I would take some time to enjoy a one-on-one text conversation or seek for actual suggestion in a particular real life scenario (hypothetically), and then I would prefer [MultiSlack] in general.''}
\end{myquote}
U23 said, 
\begin{myquote}
\textit{``For simpler questions, I would prefer that there are not many as back and forth exchanges, but I appreciate a more conversational style for questions that are more in-depth or are related to topics that I have little knowledge of.''}
\end{myquote}
U22 stated, 
\begin{myquote}
\textit{``I can say it depends on the questions asked. For the [SlackVanilla], most of the questions were straight forward and answers can be given without any follow up questions whereas [MultiSlack] provided answer based on my views and aspects.''}
\end{myquote}

% P20(2): Overall, it felt like a conversation with Assistant A as compared to transactional communication with Assistant B. I felt Assistant A personalized the responses catering to my personal taste and needs as compared to Assistant B. However, I would prefer to use Assistant B when I am looking for direct answers that are context independent.

% P25(4): It depends on what type of questions you ask. But I think the most of time, people can just ask more specifically no need the Chatbot to remind users be more specific.
\subsection{MTurk Response Analysis}
\begin{figure*}
\centering
\includegraphics[width=0.49\linewidth]{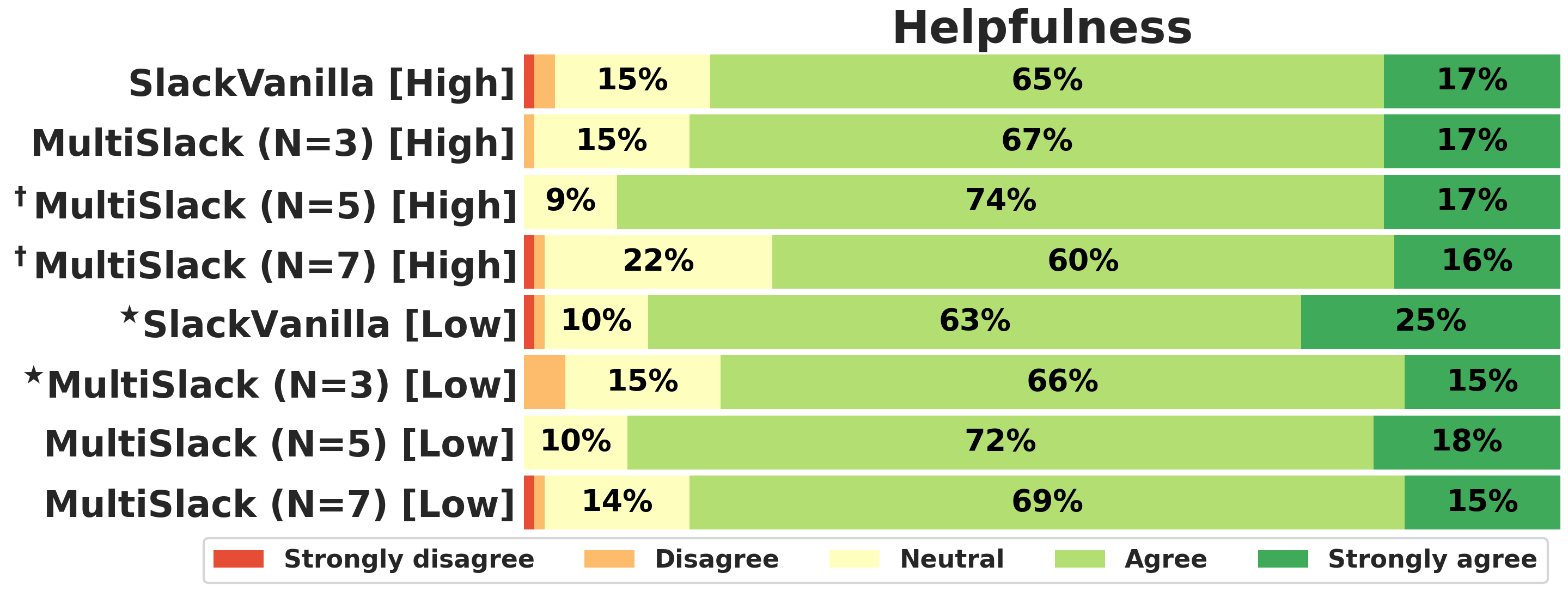}\hfil%l
\includegraphics[width=0.49\linewidth]{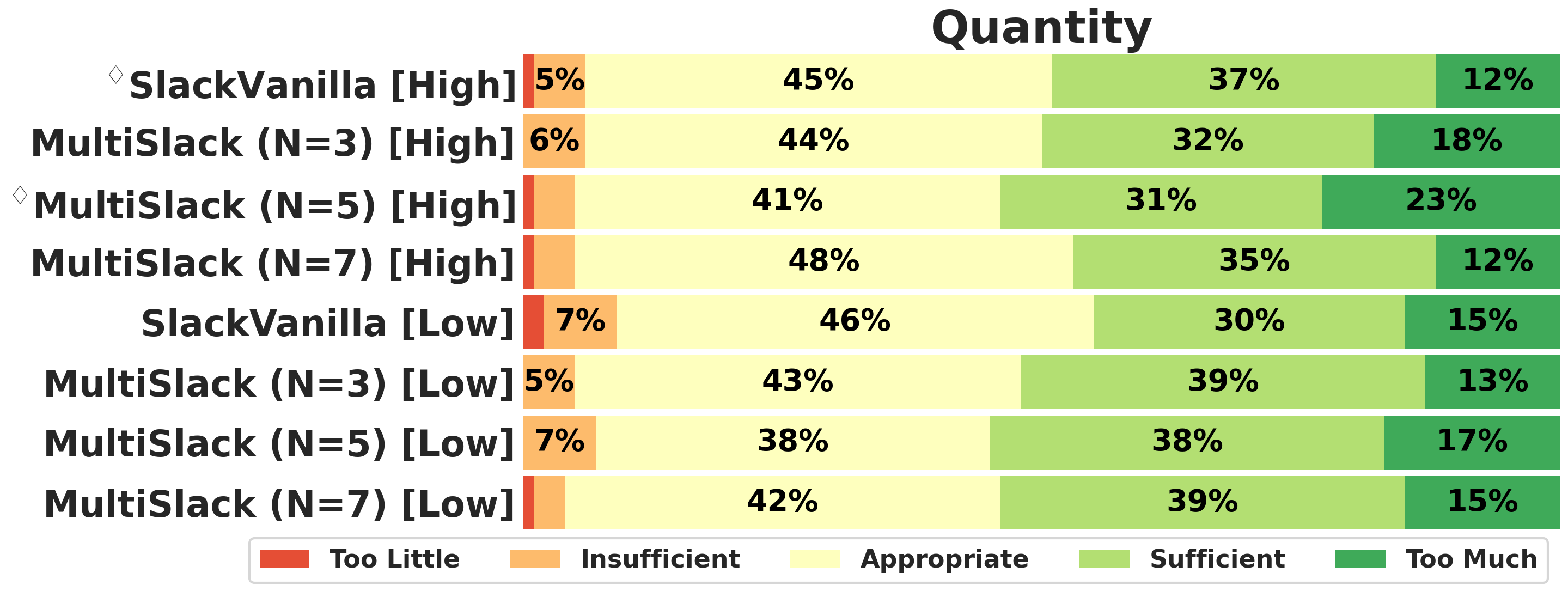}\\
\includegraphics[width=0.49\linewidth]{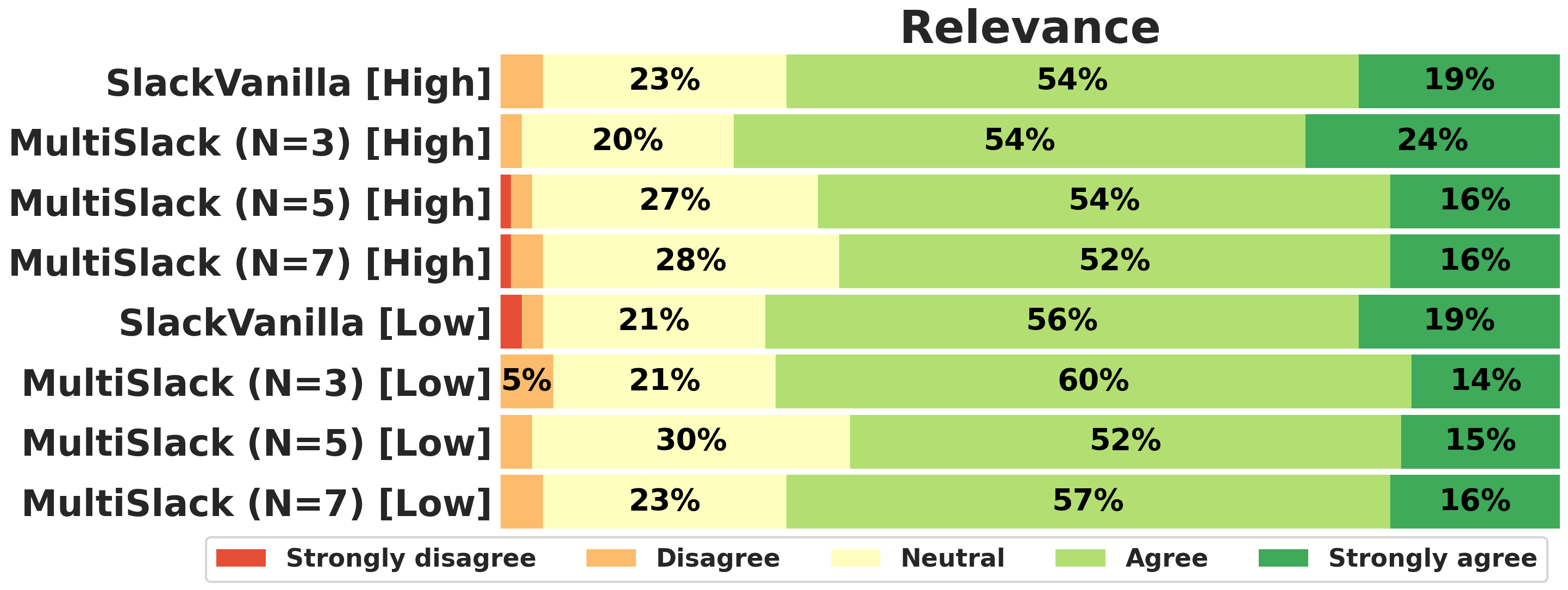}%
\hfill
\includegraphics[width=0.49\linewidth]{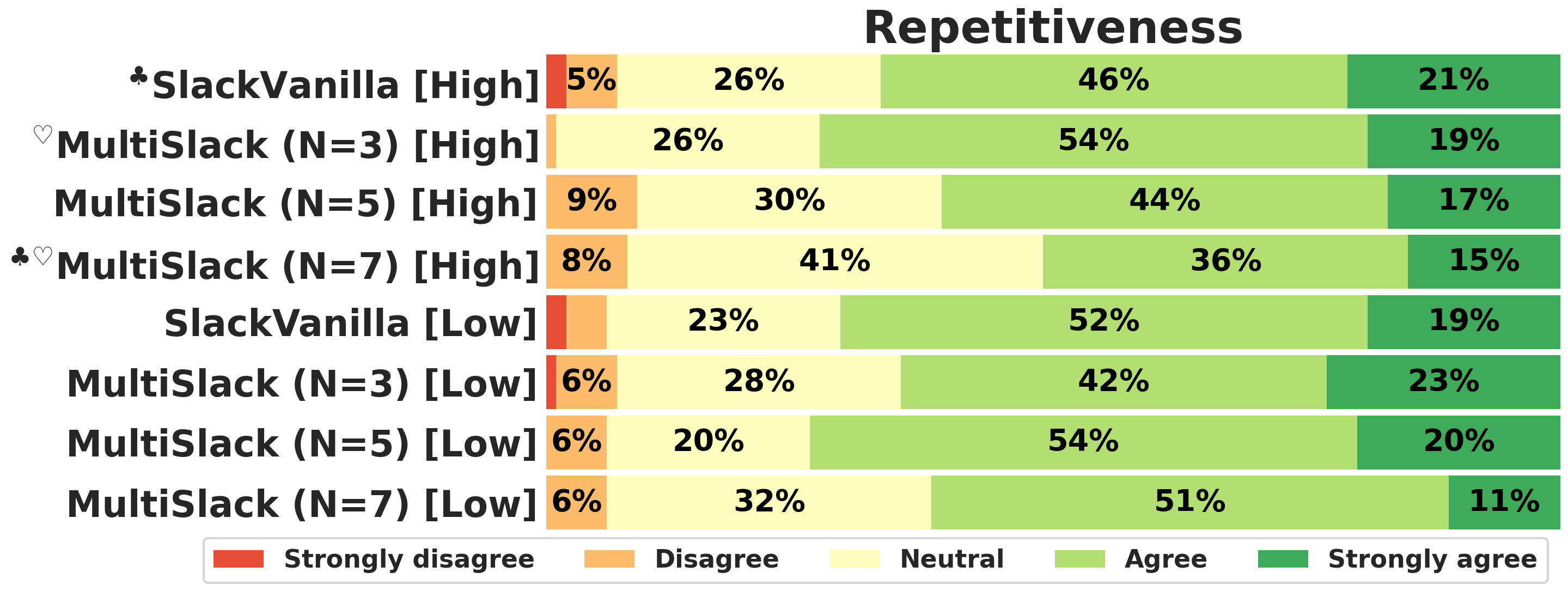}\\
\includegraphics[width=0.49\linewidth]{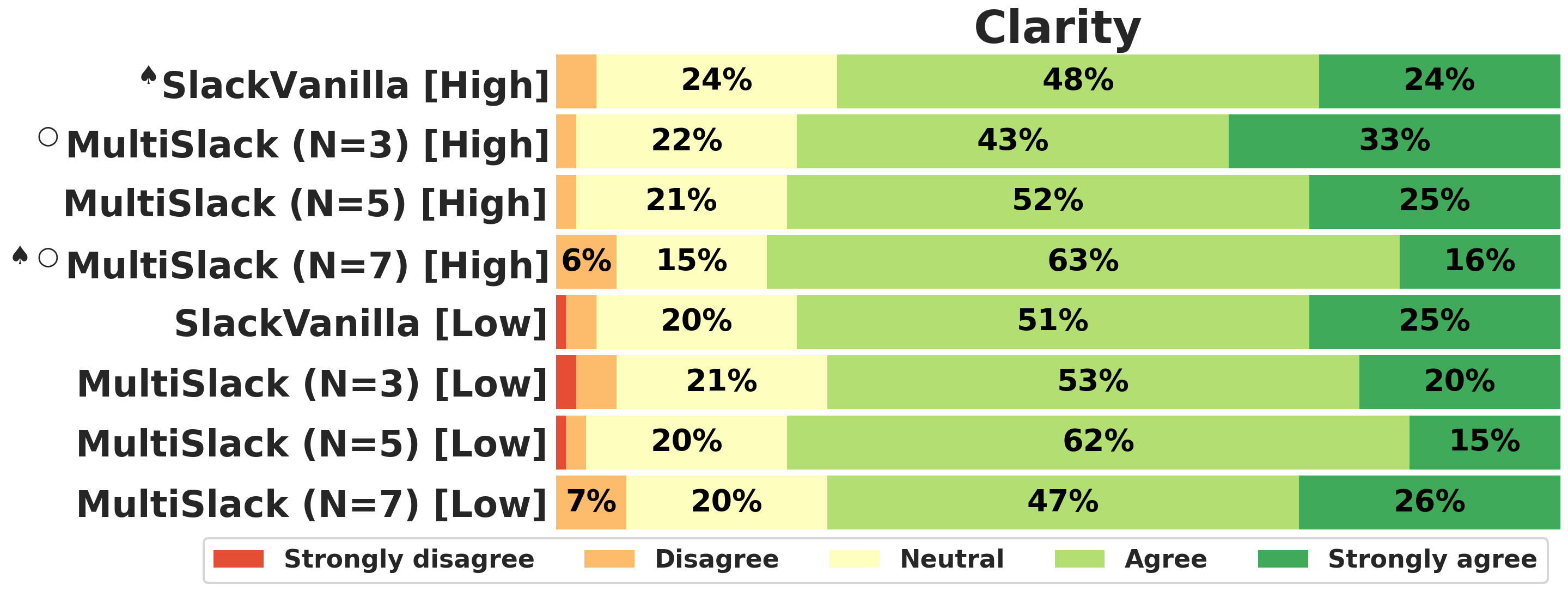}\hfill%
\includegraphics[width=0.49\linewidth]{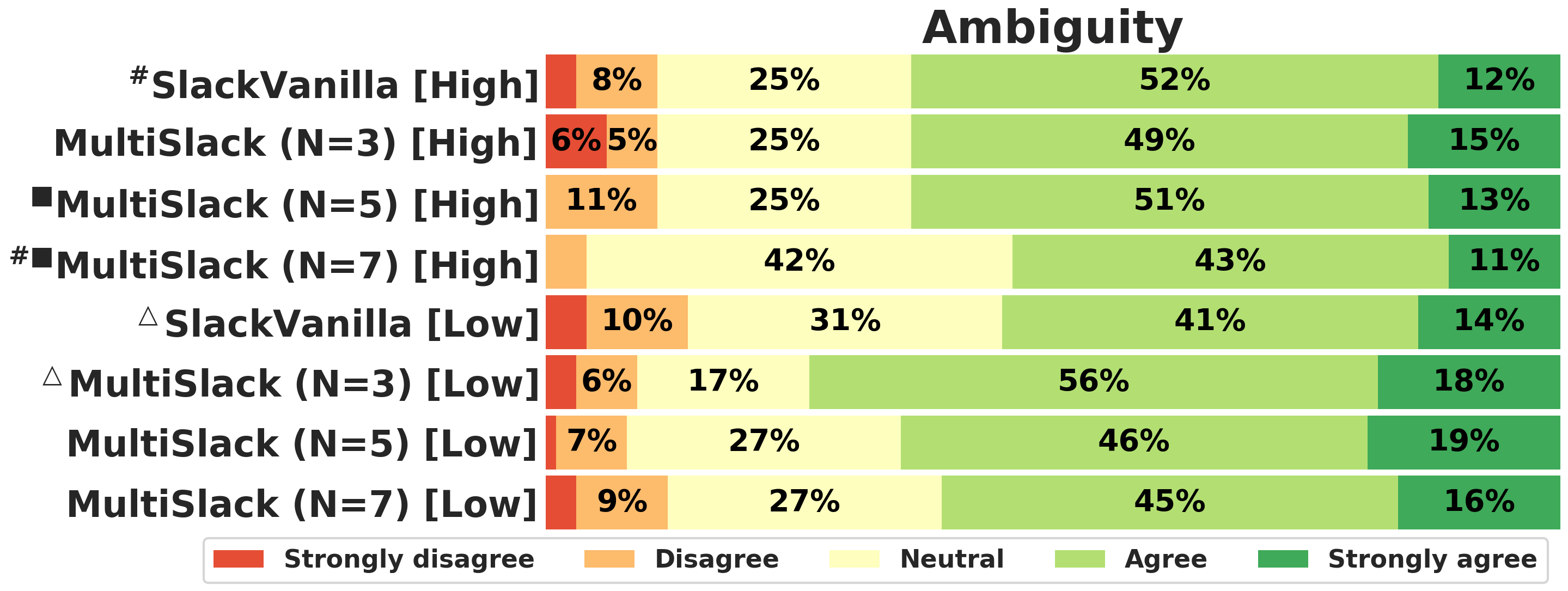}%
\caption{Rating distribution of the MTurk evaluation. Rating pairs that passed the Chi-squared Test are denoted by 
$\dagger$ (p-value = 0.046),
$\bigstar$ (p-value = 0.026),
$\diamondsuit$ (p-value = 0.041),
$\clubsuit$ (p-value = 0.013),
$\heartsuit$ (p-value = 0.01),
$\spadesuit$ (p-value = 0.022),
$\bigcirc$ (p-value = 0.015),
\# (p-value = 0.009),
$\blacksquare$ (p-value = 0.032),
and $\triangle$ (p-value = 0.005).}
    \Description{
    The figure contains six subfigures representing the distribution of the MTurk evaluation rating for the six inquiries we asked MTurk workers. Each subfigure is consisted of eight bar charts representing SlackVanilla and MultiSlack(N=3, 5, 7) for both high and low conversability questions. The bar charts are shown in five different colors to represent the five-point Likert scale, the colors used are red, orange, yellow, light green, and dark green.
    Helpfulness, located top left with red representing Strongly disagree, orange representing Disagree, yellow representing Neutral, light green representing Agree, and dark green representing Strongly agree. The eight bar charts are 
    SlackVanilla [High]: 1\% red, 2\% orange, 15\% yellow, 65\% light green, 17\% dark green.
    MultiSlack (N=3) [High]: 1\% orange, 15\% yellow, 67\% light green, 17\% dark green.
    dagger, MultiSlack (N=5) [High]: 9\% yellow, 74\% light green, 17\% dark green.
    dagger, MultiSlack (N=7) [High]: 1\% red, 1\% orange, 22\% yellow, 60\% light green, 16\% dark green.
    bigstar, SlackVanilla [Low]: 1\% red, 1\% orange, 10\% yellow, 63\% light green, 25\% dark green.
    bigstar, MultiSlack (N=3) [Low]: 4\% orange, 15\% yellow, 66\% light green, 15\% dark green.
    MultiSlack (N=5) [Low]: 10\% yellow, 72\% light green, 18\% dark green.
    MultiSlack (N=7) [Low]: 1\% red, 1\% orange, 14\% yellow, 69\% light green, 15\% dark green.
    Quantity, located top right with red representing Too little, orange representing Insufficient, yellow representing Appropriate, light green representing Sufficient, and dark green representing Too Much. The eight bar charts are 
    diamondsuit, SlackVanilla [High]: 1\% red, 5\% orange, 45\% yellow, 37\% light green, 12\% dark green.
    MultiSlack (N=3) [High]: 6\% orange, 44\% yellow, 32\% light green, 18\% dark green.
    diamondsuit, MultiSlack (N=5) [High]: 1\% red, 4\% orange, 41\% yellow, 31\% light green, 23\% dark green.
    MultiSlack (N=7) [High]: 1\% red, 4\% orange, 48\% yellow, 35\% light green, 12\% dark green.
    SlackVanilla [Low]: 2\% red, 7\% orange, 46\% yellow, 30\% light green, 15\% dark green.
    MultiSlack (N=3) [Low]: 5\% orange, 43\% yellow, 39\% light green, 13\% dark green.
    MultiSlack (N=5) [Low]: 7\% orange, 38\% yellow, 38\% light green, 17\% dark green.
    MultiSlack (N=7) [Low]: 1\% red, 3\% orange, 42\% yellow, 39\% light green, 15\% dark green.
    Relevance, located middle left with red representing Strongly disagree, orange representing Disagree, yellow representing Neutral, light green representing Agree, and dark green representing Strongly agree. The eight bar charts are 
    SlackVanilla [High]: 4\% orange, 23\% yellow, 54\% light green, 19\% dark green.
    MultiSlack (N=3) [High]: 2\% orange, 20\% yellow, 54\% light green, 24\% dark green.
    MultiSlack (N=5) [High]: 1\% red, 2\% orange, 27\% yellow, 54\% light green, 16\% dark green.
    MultiSlack (N=7) [High]: 1\% red, 3\% orange, 28\% yellow, 52\% light green, 16\% dark green.
    SlackVanilla [Low]: 2\% red, 2\% orange, 21\% yellow, 56\% light green, 19\% dark green.
    MultiSlack (N=3) [Low]: 5\% orange, 21\% yellow, 60\% light green, 14\% dark green.
    MultiSlack (N=5) [Low]: 3\% orange, 30\% yellow, 52\% light green, 15\% dark green.
    MultiSlack (N=7) [Low]: 4\% orange, 23\% yellow, 57\% light green, 16\% dark green.
    Repetitiveness, located middle right with red representing Strongly disagree, orange representing Disagree, yellow representing Neutral, light green representing Agree, and dark green representing Strongly agree. The eight bar charts are 
    clubsuit, SlackVanilla [High]: 2\% red, 5\% orange, 26\% yellow, 46\% light green, 21\% dark green.
    heartsuit, MultiSlack (N=3) [High]: 1\% orange, 26\% yellow, 54\% light green, 19\% dark green.
    MultiSlack (N=5) [High]: 9\% orange, 30\% yellow, 44\% light green, 17\% dark green.
    clubsuit, heartsuit, MultiSlack (N=7) [High]: 8\% orange, 41\% yellow, 36\% light green, 15\% dark green.
    SlackVanilla [Low]: 2\% red, 4\% orange, 23\% yellow, 52\% light green, 19\% dark green.
    MultiSlack (N=3) [Low]: 1\% red, 6\% orange, 28\% yellow, 42\% light green, 23\% dark green.
    MultiSlack (N=5) [Low]: 6\% orange, 20\% yellow, 54\% light green, 20\% dark green.
    MultiSlack (N=7) [Low]: 6\% orange, 32\% yellow, 51\% light green, 11\% dark green.
    Clarity, located bottom left with red representing Strongly disagree, orange representing Disagree, yellow representing Neutral, light green representing Agree, and dark green representing Strongly agree. The eight bar charts are 
    spadesuit, SlackVanilla [High]: 4\% orange, 24\% yellow, 48\% light green, 24\% dark green.
    bigcirc, MultiSlack (N=3) [High]: 2\% orange, 22\% yellow, 43\% light green, 33\% dark green.
    MultiSlack (N=5) [High]: 2\% orange, 21\% yellow, 52\% light green, 25\% dark green.
    spadesuit, bigcirc, MultiSlack (N=7) [High]: 6\% orange, 15\% yellow, 63\% light green, 16\% dark green.
    SlackVanilla [Low]: 1\% red, 3\% orange, 20\% yellow, 51\% light green, 25\% dark green.
    MultiSlack (N=3) [Low]: 2\% red, 4\% orange, 21\% yellow, 53\% light green, 20\% dark green.
    MultiSlack (N=5) [Low]: 1\% red, 2\% orange, 20\% yellow, 62\% light green, 15\% dark green.
    MultiSlack (N=7) [Low]: 7\% orange, 20\% yellow, 47\% light green, 26\% dark green.
    Ambiguity, located bottom right with red representing Strongly disagree, orange representing Disagree, yellow representing Neutral, light green representing Agree, and dark green representing Strongly agree. The eight bar charts are 
    \#, SlackVanilla [High]: 3\% red, 8\% orange, 25\% yellow, 52\% light green, 12\% dark green.
    MultiSlack (N=3) [High]: 6\% red, 5\% orange, 25\% yellow, 49\% light green, 15\% dark green.
    blacksquare, MultiSlack (N=5) [High]: 11\% orange, 25\% yellow, 51\% light green, 13\% dark green.
    \#, blacksquare, MultiSlack (N=7) [High]: 4\% orange, 42\% yellow, 43\% light green, 11\% dark green.
    triangle, SlackVanilla [Low]: 4\% red, 10\% orange, 31\% yellow, 41\% light green, 14\% dark green.
    triangle, MultiSlack (N=3) [Low]: 3\% red, 6\% orange, 17\% yellow, 56\% light green, 18\% dark green.
    MultiSlack (N=5) [Low]: 1\% red, 7\% orange, 27\% yellow, 46\% light green, 19\% dark green.
    MultiSlack (N=7) [Low]: 3\% red, 9\% orange, 27\% yellow, 45\% light green, 16\% dark green.
    }
    \label{fig:slackbot-turk}
\end{figure*}

% Please add the following required packages to your document preamble:
% \usepackage{booktabs}
% \usepackage{multirow}
\begin{table*}[]
\centering \small\def\arraystretch{1.15}
\begin{tabular}{@{}lllccccc@{}}
\toprule
\textbf{} & \textbf{} & \textbf{} & \textbf{\slackBaseline} & \textbf{} & \multicolumn{3}{c}{\textbf{\slackBot}} \\ \cmidrule{4-4} \cmidrule{6-8}
\textbf{} & \textbf{} & \textbf{Converasbility} & \textbf{No Conversation} & \textbf{} & \textbf{N=3} & \textbf{N=5} & \textbf{N=7} \\ \midrule
\textbf{} & \multirow{2}{*}{\textbf{\# Dialogue}} & \textbf{High} & 44 &  & 9 & 11 & 12 \\
\textbf{} &  & \textbf{Low} & 28 &  & 15 & 10 & 12 \\ \hline
\multirow{2}{*}{\textbf{Q$_1$}} & \multirow{2}{*}{\textbf{\begin{tabular}[c]{@{}l@{}}Helpfulness\\ (Strongly Disagree -\textgreater Strongly Agree)\end{tabular}}} & \textbf{High} & 3.96 &  & 4 & 4.08$^\dagger$ & 3.89$^\dagger$ \\
 &  & \textbf{Low} & 4.09$^\bigstar$ &  & 3.91$^\bigstar$ & 4.08 & 3.97 \\ \hline
\multirow{2}{*}{\textbf{Q$_2$}} & \multirow{2}{*}{\textbf{\begin{tabular}[c]{@{}l@{}}Quantity\\ (Too Little -
\textgreater Too Much)\end{tabular}}} & \textbf{High} & 3.54$^\diamondsuit$ &  & 3.6 & 3.7$^\diamondsuit$ & 3.54 \\
 &  & \textbf{Low} & 3.51 &  & 3.6 & 3.67 & 3.66 \\ \hline
\multirow{2}{*}{\textbf{Q$_3$}} & \multirow{2}{*}{\textbf{\begin{tabular}[c]{@{}l@{}}Relevance\\ (Strongly Disagree -\textgreater Strongly Agree)\end{tabular}}} & \textbf{High} & 3.87 &  & 3.99 & 3.82 & 3.81 \\
 &  & \textbf{Low} & 3.91 &  & 3.81 & 3.78 & 3.85 \\ \hline
\multirow{2}{*}{\textbf{Q$_4$}} & \multirow{2}{*}{\textbf{\begin{tabular}[c]{@{}l@{}}Repetitiveness\\ (Strongly Disagree -\textgreater Strongly Agree)\end{tabular}}} & \textbf{High} & 3.8$^\clubsuit$ &  & 3.9$^\heartsuit$ & 3.68 & 3.57$^{\clubsuit\heartsuit}$ \\
 &  & \textbf{Low} & 3.82 &  & 3.81 & 3.89 & 3.68 \\ \hline
\multirow{2}{*}{\textbf{Q$_5$}} & \multirow{2}{*}{\textbf{\begin{tabular}[c]{@{}l@{}}Clarity\\ (Strongly Disagree -\textgreater Strongly Agree)\end{tabular}}} & \textbf{High} & 3.93 &  & 4.05 & 4 & 3.91 \\
 &  & \textbf{Low} & 3.97 &  & 3.86 & 3.87 & 3.9 \\ \hline
 \multirow{2}{*}{\textbf{Q$_6$}} & \multirow{2}{*}{\textbf{\begin{tabular}[c]{@{}l@{}}Ambiguity\\ (Strongly Disagree -\textgreater Strongly Agree)\end{tabular}}} & \textbf{High} & 3.61 &  & 3.62 & 3.66 & 3.63 \\
 &  & \textbf{Low} & 3.5$^{\triangle\blacksquare}$ &  & 3.8$^\triangle$ & 3.77$^\blacksquare$ & 3.62 \\ \bottomrule
\end{tabular}
\caption{Average ratings of the MTurk evaluation. Rating pairs that passed the T-Test are denoted by 
$\dagger$ (p-value = 0.026),\\
$\bigstar$ (p-value = 0.013),
$\diamondsuit$ (p-value = 0.041),
$\clubsuit$ (p-value = 0.020),
$\heartsuit$ (p-value = 0.005),
% $\spadesuit$ (p-value = 0.022),
% $\square$ (p-value = 0.015),
% $\#$ (p-value = 0.009),
$\triangle$ (p-value = 0.004),
and $\blacksquare$ (p-value = 0.028). 
}
\label{tab:slackbot-turk-t-test}
\end{table*}
We also evaluated the quality of the conversation between the participants and both Slackbots by recruiting Amazon Mechanical Turk (MTurk) workers (the worker interface is shown in \Cref{sec:interface}). Six aspects and the accompanying inquiries of the conversations were rated on a five-point Likert scale listed below:
\begin{itemize}
    \item 
    \textbf{Helpfulness}: The assistant offer a helpful solution to the user's original question by the end of the conversation.

    \item
    \textbf{Quantity}: The assistant request an appropriate amount of information from the user, necessary for the purpose of the conversation.

    \item
    \textbf{Relevance}: The assistant seek information that is relevant to the purpose of the conversation from the user and avoid including irrelevant information.

    \item
    \textbf{Repetitiveness}: The assistant's messages are free from repeating information.

    \item
    \textbf{Clarity}: The assistant's messages are clear and easy to understand.

    \item
    \textbf{Ambiguity}: The assistant inform the user when things were unclear in the conversation and explain how it understood uncertain messages.
\end{itemize}
In this task, we asked workers to read the conversation and then rate the above six aspects.
A total of 141 conversations from 24 participants were evaluated. Notably, three conversations were excluded from the analysis among the original 144 conversations due to presence of erroneous messages.
For each task, we recruited 9 different workers with four built-in worker qualifications:
HIT Approval Rate ($\geq $98\%), Number of Approved HITs ($\geq 3000$), Local (US Only), and Adult Content.
Each task was estimated to take 70 seconds, so we set the price at \$0.2 per task (hourly wage = \$10). \Cref{tab:slackbot-turk-t-test} shows the average MTurk worker ratings on the six conversation aspects. \Cref{fig:slackbot-turk} shows the rating distribution of the six aspects.
The six aspects were inspired by Chatbot design principles proposed by ~\citet{setlur2022you}, which employed Grice’s maxims~\cite{grice1975logic}. 

% \begin{figure}
%     \centering
%     \begin{subfigure}{0.49\linewidth}
%         \includegraphics[width=\linewidth]{Figures/helpfulness.png}
%         \vspace{-6mm}
%         \caption{Helpfulness}
%         \label{fig:helpfulness}
%     \end{subfigure}
%     \begin{subfigure}{0.49\linewidth}
%         \includegraphics[width=\linewidth]{Figures/quantity.png}
%         \vspace{-6mm}
%         \caption{Quantity}
%         \label{fig:quantity}
%     \end{subfigure}
    
%     \begin{subfigure}{0.49\linewidth}
%         \includegraphics[width=\linewidth]{Figures/relevance.png}
%         \vspace{-6mm}
%         \caption{Relevance}
%         \label{fig:relevance}
%     \end{subfigure}
%     \begin{subfigure}{0.49\linewidth}
%         \includegraphics[width=\linewidth]{Figures/repetitiveness.png}
%         \vspace{-6mm}
%         \caption{Repetitiveness}
%         \label{fig:repetitiveness}
%     \end{subfigure}
    
%     \begin{subfigure}{0.49\linewidth}
%         \includegraphics[width=\linewidth]{Figures/clarity.png}
%         \vspace{-2mm}
%         \caption{Clarity}
%         \label{fig:clarity}
%     \end{subfigure}
%     \begin{subfigure}{0.49\linewidth}
%         \includegraphics[width=\linewidth]{Figures/ambiguity.png}
%         \vspace{-2mm}
%         \caption{Ambiguity}
%         \label{fig:ambiguity}
%     \end{subfigure}
%     \caption{AAAAAA}
%     \label{fig:slackbot-turk}
% \end{figure}

% The results suggested that for high-conversability questions, longer conversation length does not necessary provide more helpful solution at the end. Extending the conversation length can also resulted in repetitive information in the messages 
% and does not make the message clearer. \slackBot inform ambiguity better compare to \slackBaseline, but not necessary longer is better.
\subsubsection{Longer conversations may not be inherently better for high-conversability questions.}
The result suggested that for questions with high conversability, a longer conversation length does not necessarily yield a more helpful solution at the end. Extending the conversation length can result in repetitive information within messages, and it may not enhance message clarity. 
% While \slackBot excelled in conveying ambiguity compared to \slackBaseline, the results imply that an increase in conversation length does not consistently equate to improved outcomes. 
While \slackBot may potentially address ambiguity better compared to \slackBaseline regardless of a question's conversability, fluctuations in ambiguity ratings imply that an increase in conversation length does not consistently equate to improved outcomes. In essence, the MTurk evaluation suggests that longer interactions are not inherently superior in addressing high-conversability questions.

% \begin{table}[]
% \begin{tabular}{|l|c|c|}
% \hline
% Q4   & Assistant A & Assistant B \\ \hline
%      & 3.11        &             \\ \cline{2-2}
% High & 2.71        & 3.55        \\ \cline{2-2}
%      & 3.75        &             \\ \hline
%      & 3.33        &             \\ \cline{2-2}
% Low  & 3.6         & 3.25        \\ \cline{2-2}
%      & 3.42        &             \\ \hline
% \end{tabular}
% \end{table}

% \begin{table}[]
% \begin{tabular}{|l|c|c|}
% \hline
% Q5   & Assistant A & Assistant B \\ \hline
%      & 3.44        &             \\ \cline{2-2}
% High & 3.86        & 3.57        \\ \cline{2-2}
%      & 4.25        &             \\ \hline
%      & 4.07        &             \\ \cline{2-2}
% Low  & 4.2         & 4.39        \\ \cline{2-2}
%      & 4.17        &             \\ \hline
% \end{tabular}
% \end{table}

% \begin{table}[]
% \begin{tabular}{|l|c|c|}
% \hline
% Q6   & Assistant A & Assistant B \\ \hline
%      & 3.44        &             \\ \cline{2-2}
% High & 3.57        & 3.64        \\ \cline{2-2}
%      & 4.33        &             \\ \hline
%      & 4.13        &             \\ \cline{2-2}
% Low  & 4.2         & 4.36        \\ \cline{2-2}
%      & 4.17        &             \\ \hline
% \end{tabular}
% \end{table}

% \subsubsection{Study Procedure}

% \subsubsection{Findings}
% study1: under different situation, prefer different modality
% study2: different type of conversation/questions, different conversation style can be better

%\input{Sections/4-study-procedure}

%\section{Findings}
%\input{Sections/5-results}
%\section{Experimental Results}
%\input{Sections/4-results}

\section{Discussion}

% This paper introduces two case studies advocating integrating situation awareness in conversational assistants. 
% Firstly, the Patrol Study demonstrates that users adjust their communication methods according to their situations, validating that assistants can communicate more effectively by recognizing and adapting to these situations.
% Secondly, the Conversation Length Study shows that assistants enhance user experiences by adapting conversation formats to specific situations. 

% The concept of ``situation'' warrants clarification. 
% %A relevant question is what is the scope of ``situation''.
% While ``situation'' and ``context'' can have varied meanings across different domains, like ubiquitous computing, affective computing, and NLP/dialogue systems, this work intentionally separates conversation content and format. 
% We focus on adjusting the conversation format using factors external to content, countering the mainstream dialogue system research's emphasis on content, and underscoring the importance of meta aspects. 
% Enabling extensive, human-like conversations involves more than merely disseminating information and crafting fluent responses.

This paper presents a case study examining the influence of message length in interactions with a GPT-4-powered Slack chatbot. Our observations suggest that users tend to engage more deeply in discussions involving questions with high conversational potential. This is reflected in the increase in self-reported satisfaction and perceived helpfulness scores. However, our MTurk assessment reveals a more nuanced picture (see \Cref{tab:slackbot-turk-t-test} and \Cref{fig:slackbot-turk}): while helpfulness scores do improve as the number of conversational turns ($N$) increases for these high-conversational questions, there is a notable decline when $N$ reaches seven. Additionally, other questions evaluating the assistant's performance do not show a corresponding improvement with an increase in $N$. This suggests that beyond a certain point, more conversation does not necessarily equate to higher effectiveness.

% This paper introduces a case study discussing the impact of length on conversation with a GPT-4 backed Slack chatbot.
% We observed that users might want to get involved in a discussion for high-conversatbility questions as evidenced by the raise of the self-reported satisfication score and helpfulness score.
% However, the Mturk evaluation shows that the helpfulness does not essentially increase (for high-conversability questions, helpfulness increases when N increases but significantly drops when N=7). Other ratings on the performance of the assistant also suggest that assistant does not perform better when N increases. 

The effectiveness of the prompts used in \slackBaseline and \slackBot might be a factor in generating suboptimal responses. Merely instructing a large language model (LLM) to pose follow-up questions does not guarantee the generation of high-quality inquiries that meaningfully advance the conversation towards an effective resolution. It is also conceivable that the LLM lacks the capability to formulate truly pertinent questions, which requires a deep understanding of the user's overall situation, condition, or objectives. This limitation is significant, given users’ general unfamiliarity with the full extent of the LLM's capabilities, whether robust or limited.

Regarding the ideal conversational length, denoted as $N$, our findings suggest that it varies unpredictably across different topics, even among those with high conversational potential. There is not a linear or straightforward trend indicating that longer or shorter conversations are inherently better. Instead, the ideal length $N$ appears to be dynamic, influenced by the nature and quality of follow-up questions asked as the conversation unfolds. Poor quality inquiries by the bot may necessitate a longer conversation (increasing $N$), whereas incisive, progress-driving questions might shorten the required conversational length (decreasing $N$). Thus, $N$ is not a fixed value but is fluid and context-dependent, shifting in response to the evolving quality of the dialogue.

% It is possible that the prompt used for \slackBaseline and \slackBot is not good enough, resulting in a bad utterance. Simply prompting by asking LLM to ask follow up questions does not translate into LLM asking high quality questions that actually push the conversation further towards a good solution. On the contrary, it is also possible that LLM does not have the capability to ask truly relevant questions, not just related, but fully understand the overall situation/condition/attempt of the user. We as users do not know it's capability, regardless of whether it is good or bad.

% The ideal length $N$ for different topics can be random, even for high-conversatbility questions. not necessary linear or have straight forward tendency (not longer better nor shorter better).
% The ideal length $N$ can be dynamic and related to what follow up questions are asked as the conversation is progressing, instead of a set amount before the conversation started. If the bot inquiries are bad and does not contribute, $N$ can increase. Vice versa, if key inquiries are made and pushed the conversation forward, $N$ can decrease.

\paragraph{Limitations.}
We acknowledge several limitations in this study. 
Firstly, the research concentrates on one specific aspect of conversation: conversation length, leaving additional attributes, such as output modality and delivery channel, unexplored.
Secondly, the study's limited scale and bias towards younger and well-educated users with prior experience interacting with LLMs challenge the generalizability of the results, necessitating further investigation.
% Thirdly, while we advocate for conversational assistants to adjust format based on the situation, the alteration of format, such as conversation length dictating word limits and distinct error types in voice and text inputs, inevitably influences the content, blurring the boundary between format and content.
Lastly, we recognize the potential trade-offs between safeguarding user privacy and meeting their needs. Our approach, requiring conversational assistants to be situationally aware, sacrifices some user privacy for convenience, potentially introducing additional risks. This topic warrants further exploration in future research.

\section{Conclusion and Future Work}
This paper introduces a case study investigating the impact of conversation length on users' conversation with LLM supported conversational assistants.
It demonstrates that conversational assistants, which strategically adjust conversation formats to current situations, confer user benefits.
This study argues for the development of future conversational assistants that can recognize user situations and accordingly adjust conversational formats.
Looking ahead, we aim to explore the potential for conversational assistants to automatically detect situations and make strategic communication decisions. 
Just as we do not explicitly instruct our friends on conversational preferences, we should not need to configure conversational assistants during each interaction.
Leveraging the enhanced AI capabilities of LLMs to address HCI challenges is crucial for facilitating human-like, meaningful conversations that transcend mere information provision and fluent response generation.

\begin{acks}
We are grateful to the anonymous reviewers for their constructive feedback, and to the participants and MTurk workers for their invaluable contributions to our study.
\end{acks}
%%
%% The acknowledgments section is defined using the "acks" environment
%% (and NOT an unnumbered section). This ensures the proper
%% identification of the section in the article metadata, and the
%% consistent spelling of the heading.
%\begin{acks}
%To Robert, for the bagels and explaining CMYK and color spaces.
%\end{acks}

%%
%% The next two lines define the bibliography style to be used, and
%% the bibliography file.
%\bibliographystyle{ACM-Reference-Format}
%\bibliography{sample-base}
\bibliographystyle{Style/ACM-Reference-Format}
% \bibliography{Bibtex/main}
%%% -*-BibTeX-*-
%%% Do NOT edit. File created by BibTeX with style
%%% ACM-Reference-Format-Journals [18-Jan-2012].

\newpage

%%
%% If your work has an appendix, this is the place to put it.
\appendix

\section{Question List for the Study}
\label{sec:question-pool}
\Cref{tab:question-pool-high} displays the 20 selected questions with highest Conversation scores, \Cref{tab:question-pool-low} presents the 20 selected questions with lowest Conversation scores for our study.
\begin{table*}[h]
    \centering \small\def\arraystretch{1.35}
    \begin{tabular}{@{}cp{14cm}@{}}
        \toprule
      {\textbf{Conversation Score}} & \multicolumn{1}{c@{}}{\textbf{Question}} \\ \midrule
        4.33                                  &What happens after we die? \\ \hline
        4.33                      &What to do when you are feeling lost in life?  \\ \hline
        4.33                            &Recommendation for a vacuum cleaner?  \\ \hline
        4.22                                &Is it possible to make a good situation out of any bad situation?  \\ \hline
        4.22                        &What happens when you chew a poisonous flower for a few seconds but spit it out? \\ \hline
        4.22                 &Is politics more entertaining now than it was in decades prior? \\ \hline
        4.22	              &What is a fine tuned universe? Why is gravity fine-tuned? \\ \hline
        4.22                 &What do you do when you can’t get an anxiety-inducing thought out of your head? \\ \hline
        4.22                      &how much pain did you feel after wisdom tooth removal? \\ \hline
        4.22                       &How to control emotions? How do people control their emotions when they lost their loved one?  \\ \hline
        4.22                      &This housing market is wild. Is it going to last the next 4 years?  \\ \hline
        4.11                               &Why have humans gotten so weak physically?Like a few hundred years ago,people were cut up by swords and literally lost limbs but still were able to fight,but now we scared of things like needles. \\ \hline
        4.11             &What do I watch? I just downloaded Reddit and Netflix. What do I watch?  \\ \hline
        4.11                        &My friend is meeting her boyfriend’s family this weekend over dinner. She is not Mormon, but his family is. He has separated himself from the church, but his family has not, though they are all on good terms. What are appropriate topics of conversation?  \\ \hline
        4.11                               &What's the difference between an alpha male and a sigma male? \\ \hline
        4.11                               &Can someone explain the fastest way to learn Pokemon so I can understand what the hell my kids are talking about please? \\ \hline
        4.11                               &How to say to an Asian parent that you like to move out \\ \hline
        4.11                               &How do you approach someone you used to be friends with without making it awkward? \\ \hline
        4.11                               &I have LG UX 250 sound system but I can't switch to AUX option since I lost original remote. Anymote app doesn't work, so it's there any other solution? \\ \hline
        4.11                               &``The more you have of something, the less you value it, and the more you take it for granted''. What is an example of this in our day to day life? \\ \bottomrule
        \end{tabular}
    \caption{20 highest Conversation score questions selected for the study.}
    \label{tab:question-pool-high}
\end{table*}

\begin{table*}[h]
    \centering \small\def\arraystretch{1.35}
    \begin{tabular}{@{}cp{14cm}@{}}
        \toprule
        {\textbf{Conversation Score}} & \multicolumn{1}{c@{}}{\textbf{Question}} \\ \midrule
        1.89                               &What is a newsstand? \\ \hline
        2                               &What flavor is Mountain Dew Baja blast? \\ \hline
        2.33                               &What are some super low luminous e6 bulbs I can find for a chandelier? \\ \hline
        2.33                               &What are some incurable brain diseases that can affect people younger than 20? What is the life expectancy? \\ \hline
        2.44                               &I tried to delete all contents of a dvd but there is a protection that doesn't, anyone know how to deal with that? \\ \hline
        2.44                               &When a male ejaculates does the sperm come from just one testicle or is it like a 50/50 distribution? \\ \hline
        2.44                               &How do you report problems related to windows to microsoft and get fix to that problem? \\ \hline
        2.56                               &What is called when food is more expensive than other materials (like electronics) and otherwise? It is related to inflation and deflation, forgot what’s it’s the name of that situation. \\ \hline
        2.56                               &Is it normal for fingers to turn blue when taking blood pressure with automatic machine? \\ \hline
        2.56                               &Why is curry sososo spicy? \\ \hline
        2.56                               &What word or phrase could replace ``trump card''? \\ \hline
        2.56                               &Cup cooler / warmer mug search: Looking for a mug that is insulated on the outside, but metal-to-beverage on the bottom. \\ \hline
        2.56                               &do you know any website to watch marvel what if? \\ \hline
        2.56                               &How does 2+2=5? \\ \hline
        2.56                               &There's a game similar to space invaders with bug like aliens and it had boss battles from around 10 years ago and it had a 2 level free trial before paying for it. Does anyone know the name of the game? \\ \hline
        2.56                               &what are some thrifting apps that you can use globally, other than depop? \\ \hline
        2.56                               &My Aunt lend me one hundred dollar bill to play with her at casino and I won 6k she gave me two hundred back for winning how much will IRS charge me? I live in Arizona \\ \hline
        2.67                               &Does anyone know a good recipe for gumbo soup \\ \hline
        2.67                               &How do I do a free reverse phone look up? \\ \hline
        2.67                               &Does Eminem sing the Vocals of the Queen sample he used for Beautiful? \\ \bottomrule
        \end{tabular}
    \caption{20 lowest Conversation score questions selected for the study.}
    \label{tab:question-pool-low}
\end{table*}

% philosophy	What happens after we die ?
% life_suggestion	What to do when you are feeling lost in life?
% work_place	What is the best way to subtly and consistently annoy your coworkers, without them ever realising it's your fault?
% mental_health	What do you do when you can’t get an anxiety-inducing thought out of your head?
% life_suggestion	How to control emotions? How do people control their emotions when they lost their loved one?
% life_suggestion	What steps should I take towards moving out of my parents house? I'm at the ripe old age of 16 when the state of Pennsylvania graciously gives me the chance to operate a motor vehicle. What can I do to get myself headed in the direction of living on my own?
% mental_health	Let me start off by saying, yes I've tried most of the normal avenues, and yet my mind is still filled with thoughts of nihilism. Every moment of my life feels like I'm just waiting. Not for anything in particular, just something. Is there anywhere for people like me to go, and just disappear?
% history_old_days_future	Is politics more entertaining now than it was in decades prior?
% life_suggestion	Is it possible to make a good situation out of any bad situation?
% medical_health_diet	What happens when you chew a poisonous flower for a few seconds but spit it out?
% medical_health_diet	how much pain did you feel after wisdom tooth removal?
% economy	This housing market is wild. Is it going to last the next 4 years?
% science	What is a fine tuned universe? Why is gravity fine-tuned?
% medical_health_diet	How to reduce one sided cheek fat?

\section{Implementation Details of the two SlackBots}
\label{sec:appendix-slackbot-implementation}
Both Slackbots, \slackBaseline and \slackBot, are initiated with the prompt:
\begin{myquote}
    \textit{You are a helpful assistant that helps users answer their questions.}
\end{myquote}
\slackBot is additionally set to ask a certain amount of follow-up questions before providing an answer.
Such a constraint is imposed by adding the following instruction to the end of the users' message after users asked their initial question. \textbf{N} is used to denote the amount of follow-up questions required in the example below.
\begin{myquote}
    \textit{Please answer the question ``user message'' that suits my situation best by asking me follow up questions. You HAVE TO ask me \textbf{[N]} follow up questions. Do not ask me all the questions at once. Progress base on the content of our conversation and ask questions based on my previous responses. Ask the first follow up question and wait for my response. Prepare the second follow up question according to my response to your first question. Same process goes for the remaining follow up questions. Do not ask me the same follow up questions. After my response to all the follow up questions, conclude the information I gave and provide an answer to the original question. Make my interaction with you like conversing with a real human agent, do not let me know how much follow up questions your are going to ask me and do not make the conversation bullet point oriented.  Gather information from the follow up question in a subtle way and do not ask the questions through the ways of ``Here is my \# question?", ``1. ...'' etc.}
\end{myquote}
To ensure that the assistant strictly adheres to the specified follow-up question count, we include the following instruction at the end of each user's response. \textbf{N} is used to denote the amount of follow-up questions required in the example below.
\begin{myquote}
    \textit{You still have to ask me \textbf{[N]} more follow up questions. Do not ask me all the questions at once. Progress base on the content of our conversation and ask questions based on my previous responses. Do not ask me the same follow up questions.}
\end{myquote}
To signal to the assistant when it should stop asking questions and provide an answer, the following instruction is added to the final message sent by the participants:
\begin{myquote}
    \textit{Do not ask me anymore follow up questions. Conclude the information I gave and provide an answer to the original question by impyling that there will be no more follow questions and is concluding the original question asked. When you conclude, please make it explicit so I know you are concluding the question. Maybe say something similar but not limited to ``To answer your original question...'' etc}
\end{myquote}

\slackBot gathers information throughout its conversation with users, and the questions it asks are based on users' previous responses. The final answer the assistant provides is also derived from the entirety of the conversation.

\paragraph{History for \slackBot configuration.}
\begin{choruschat}
    \chatgptSystem{You are a helpful assistant that helps users answer their questions by asking a certain amount of follow up questions and give a conclusive answer in the end.}
    \chatgptUser{Please answer the question ``How to actively socialise and make friends? (coming from someone who turned from a extrovert to an introvert in a span of two years.)'' that suits my situation best by asking me follow up questions. You HAVE TO ask me THREE follow up questions. Ask the first follow up question and wait for my response. Prepare the second follow up question according to my response to your first question. Same process goes for the remaining follow up questions. After my response to all the follow up questions, conclude the information I gave and provide an answer to the original question. Make my interaction with you like conversing with a real human agent, do not let me know how much follow up questions your are going to ask me and do not make the conversation bullet point oriented.  Gather information from the follow up question in a subtle way and do not ask the questions through the ways of ``Here is my \# question?'', ``1. ...'' etc.}
    \chatgptAssistant{Sure, I can help you with that! What initially caused the shift from being an extrovert to becoming more introverted over the past two years?}
    \chatgptUser{I can't recall. You still have to ask me two follow up questions. Do not ask me all the questions at once. Progress base on the content of our conversation and ask questions based on my previous responses.}
    \chatgptAssistant{That's alright. In that case, could you tell me a bit about your current socializing habits? How often do you engage in social activities or spend time with others?"}
    \chatgptUser{I spend a lot of time home by myself or meeting with small groups if any.}
    \chatgptAssistant{Thank you for sharing that. Now, let's talk about your interests and hobbies. What are some activities or hobbies that you enjoy doing in your free time?}
    \chatgptUser{I watch TV and listen to music. This will be the last follow up question you ask me. Do not ask me anymore follow up questions and conclude the information I gave and provide an answer to the original question.}
    \chatgptAssistant{Thank you for sharing that. Based on the information you provided, here are some suggestions on how to actively socialize and make friends, considering your shift from being an extrovert to an introvert: Explore your interests in a social context: Look for local events or communities centered around your interests, such as music concerts, TV show or movie clubs, or even online forums and discussion groups. This way, you can connect with like-minded individuals who share your passion for the things you enjoy. Start with small group activities: If you prefer smaller gatherings, consider joining or organizing activities that involve a few people. This can include game nights, movie marathons, or even casual meetups at a coffee shop. Starting with smaller groups can help ease you back into socializing and create a more comfortable environment. Volunteer or join social groups: Engaging in activities where you can contribute to a cause or join a community can be a great way to meet new people. Consider volunteering for a cause you care about, joining a book club, or participating in a local sports team. These activities provide a shared purpose and can help you connect with others who have similar values or interests. Remember, building new friendships takes time and effort. Be patient with yourself and take small steps towards socializing. Don't be afraid to initiate conversations or reach out to people who share common interests. And most importantly, be open to new experiences and be yourself. Good luck on your journey to actively socialize and make new friends!}
    \chatgptSystem{You are a helpful assistant that helps users answer their questions by asking a certain amount of follow up questions and give a conclusive answer in the end.},
    \chatgptUser{Please answer the question ``Should I get a cat or a dog'' that suits my situation best by asking me follow up questions. You HAVE TO ask me FIVE follow up questions. Ask the first follow up question and wait for my response. Prepare the second follow up question according to my response to your first question. Same process goes for the remaining follow up questions. After my response to all the follow up questions, conclude the information I gave and provide an answer to the original question. Make my interaction with you like conversing with a real human agent, do not let me know how much follow up questions your are going to ask me and do not make the conversation bullet point oriented.  Gather information from the follow up question in a subtle way and do not ask the questions through the ways of ``Here is my \# question?'' etc.}
    \chatgptAssistant{Sure! I'd be happy to help you decide whether getting a cat or a dog would be the best choice for you. May I ask if you have any experience owning either a cat or a dog?"}
    \chatgptUser{no this will be my first pet of any kind. You still have to ask me four follow up questions. Do not ask me all the questions at once. Progress base on the content of our conversation and ask questions based on my previous responses.}
    \chatgptAssistant{Got it! That's helpful to know. Can you tell me a bit about your living situation? Do you live in a house or an apartment?"}
    \chatgptUser{I live in an apartment with a 9 to 5 job. You still have to ask me three follow up questions. Do not ask me all the questions at once. Progress base on the content of our conversation and ask questions based on my previous responses.}
    \chatgptAssistant{Understood! Can you also tell me about your daily routine? Are you typically home in the evenings, or do you have extracurricular activities that take up a lot of your time?"}
    \chatgptUser{I am mostly home after work, not much going on. You still have to ask me two follow up questions. Do not ask me all the questions at once. Progress base on the content of our conversation and ask questions based on my previous responses.}
    \chatgptAssistant{Thank you for that information. How much time are you willing to dedicate to taking care of a pet? Do you have a preference for a pet that requires more or less attention?}
    \chatgptUser{i am likely exhausted after work so probably a pet that requires less attention}
    \chatgptAssistant{I understand. Last question, do you have any allergies or concerns related to pet hair or dander?}
    \chatgptUser{I don't think i am allergic to hair but would like to play safe. This will be the last follow up question you ask me. Do not ask me anymore follow up questions and conclude the information I gave and provide an answer to the original question.}
    \chatgptAssistant{Thanks for letting me know! Based on the information you've provided, a cat may be a better choice for you. While both cats and dogs require care and attention, cats tend to be more independent and require less maintenance than dogs do. They also don't need as much space to exercise, making them a good choice for apartment living. Additionally, cats are generally less demanding of attention and do quite well on their own during the day while their owners are away at work. They require less exercise and don't need to be taken outside to go to the bathroom, which can make them a more convenient option for someone with a busy schedule. In terms of allergies, while no pet is entirely hypoallergenic, cats tend to produce less dander and shed less than dogs, making them a better choice for those with allergies or concerns about pet hair. I hope this information helps you make an informed decision!}    
\end{choruschat}

\section{Questionnaires for the User Study}\label{sec:survey}
\Cref{tab:study2-survey-A} and \Cref{tab:study2-survey-B} show the questionnaire questions we asked in \Cref{sec:study-2}.
% Please add the following required packages to your document preamble:
% \usepackage{booktabs}
\begin{table*}
\small
\centering\def\arraystretch{1.15}
\begin{tabular}{@{}lll@{}}
\toprule
\textbf{\#} & \textbf{Aspect} & \textbf{Questionnaire Question} \\ \midrule
Q1 & Reach-Out & \begin{tabular}[c]{@{}l@{}}If I have this question, I would reach out to other people, such as friends, family members, colleagues,\\ or experts, to get help. As compared to finding the answers by looking up information by myself with\\ a computer or a smartphone.\\(1) Strongly Disagree  (2) Disagree  (3) Neutral  (4) Agree  (5) Strongly Agree \end{tabular} \\ \hline
Q2 & Conversability & \begin{tabular}[c]{@{}l@{}}If I have this question, I would prefer to have a conversation regarding the details of the question and\\ have a further discussion with the answerer. As compared to asking the question as is and waiting for\\ the answers.\\(1) Strongly Disagree 
 (2) Disagree  (3) Neutral  (4) Agree  (5) Strongly Agree\end{tabular} \\ \hline
Q3 & Contribution & \begin{tabular}[c]{@{}l@{}}Follow-up questions the assistant asked contribute to providing an answer well-suited for your situation.\\(1) Strongly Disagree  (2) Disagree  (3) Neutral  (4) Agree  (5) Strongly Agree\end{tabular} \\ \hline
Q$_A$ & Enough Count & \begin{tabular}[c]{@{}l@{}}With the goal of answering your question sufficiently, how would you rate the number of follow-up\\ questions the assistant asked before providing its answer?\\(1) Not Enough 
  --->  (3) Neutral   --->  (5) Too Many\end{tabular} \\ \hline
Q$_B$ & Helpfulness & \begin{tabular}[c]{@{}l@{}}At the end of the conversation, the assistant provided a helpful solution to the question you initially asked.\\(1) Strongly Disagree  (2) Disagree  (3) Neutral  (4) Agree  (5) Strongly Agree\end{tabular} \\ \hline
Q$_C$ & Satisfaction & \begin{tabular}[c]{@{}l@{}}Overall, are you satisfied with how the assistant help you solve the question?\\(1) Very Dissatisfied  (2) Dissatisfied  (3) Neutral  (4) Satisfied  (5) Very Satisfied\end{tabular}
\\ \bottomrule
\end{tabular}
\caption{Questionnaire questions used for \slackBot.}
\label{tab:study2-survey-A}\vspace*{-6pt}
% \vspace{-5mm}
\end{table*}

\begin{table*}
\small
\centering\def\arraystretch{1.15}
\begin{tabular}{@{}lll@{}}
\toprule
\textbf{\#} & \textbf{Aspect} & \textbf{Questionnaire Question} \\ \midrule
Q1 & Reach-Out & \begin{tabular}[c]{@{}l@{}}If I have this question, I would reach out to other people, such as friends, family members, colleagues,\\ or experts, to get help. As compared to finding the answers by looking up information by myself with\\ a computer or a smartphone.\\(1) Strongly Disagree  (2) Disagree  (3) Neutral  (4) Agree  (5) Strongly Agree \end{tabular} \\ \hline
Q2 & Conversability & \begin{tabular}[c]{@{}l@{}}If I have this question, I would prefer to have a conversation regarding the details of the question and\\ have a further discussion with the answerer. As compared to asking the question as is and waiting for\\ the answers.\\(1) Strongly Disagree 
 (2) Disagree  (3) Neutral  (4) Agree  (5) Strongly Agree\end{tabular} \\ \hline
Q$_A$ & Should Ask More & \begin{tabular}[c]{@{}l@{}}With the goal of answering your question sufficiently, do you think the assistant should ask more follow-up \\questions before providing its answer?\\(1) Strongly Disagree  (2) Disagree  (3) Neutral  (4) Agree  (5) Strongly Agree\end{tabular} \\ \hline
Q$_B$ & Helpfulness & \begin{tabular}[c]{@{}l@{}}At the end of the conversation, the assistant provided a helpful solution to the question you initially asked.\\(1) Strongly Disagree  (2) Disagree  (3) Neutral  (4) Agree  (5) Strongly Agree\end{tabular} \\ \hline
Q$_C$ & Satisfaction & \begin{tabular}[c]{@{}l@{}}Overall, are you satisfied with how the assistant help you solve the question?\\(1) Very Dissatisfied  (2) Dissatisfied  (3) Neutral  (4) Satisfied  (5) Very Satisfied\end{tabular}
\\ \bottomrule
\end{tabular}
\caption{Questionnaire questions used for \slackBaseline.}
\label{tab:study2-survey-B}\vspace*{-6pt}
% \vspace{-5mm}
\end{table*}

\section{Example Conversations for the Study}
\label{sec:example-conversation-2}
We show four complete Slack conversations on the same question \textit{``This housing market is wild. Is it going to last the next 4 years?''} between participants and the four Slackbot configurations (\slackBaseline, \slackBot with $N=3$, $N=5$, and $N=7$) 

\paragraph{Conversation between U25 and \slackBaseline.}
\begin{choruschat}
    \chatgptUser{This housing market is wild. Is it going to last the next 4 years?}
    \assistantB{As an AI, I don't predict future events. However, I can tell you that many factors influence the housing market, including interest rates, the economy, and supply and demand. Many experts have varying opinions on where the housing market will go. It's always a good idea to consult with a real estate professional or financial advisor for personalized advice based on your individual circumstances.}
\end{choruschat}

\paragraph{Conversation between U23 and \slackBot $N=3$.}
\begin{choruschat}
    \chatgptUser{This housing market is wild. Is it going to last the next 4 years?}
    \assistantA{Absolutely, I can assist with that! To better understand your situation, could you tell me if you're currently a homeowner, renting, or looking to buy a property?}
    \chatgptUser{I'm renting and looking to buy a property}
    \assistantA{Understood. Are you looking to buy a property for living or for investment purposes?}
    \chatgptUser{I'm unsure yet, perhaps both}
    \assistantA{Sure, I understand how the decision could be affected by various factors. For the last question, what is your priority in buying the property? Are you more focused on the location, price, or the potential for property value appreciation?}
    \chatgptUser{I don't really have a clear idea and I think I would need more help deciding}
    \assistantA{To answer your original question, predicting the housing market is challenging even for experts. Various factors such as economic health, interest rates, and government policies can influence the real estate market. Currently, the market is indeed quite heated, but it's uncertain whether it will remain this way for the next four years.\\
    Based on your responses, since you are currently renting and considering buying a property, either for living or investment purposes, it's important to carefully consider your financial situation and your long-term plans. Given the volatile nature of the market, it can be beneficial to consult with a real estate advisor or financial planner.\\
    One factor to consider is the historically low mortgage rates which make borrowing more affordable, though they've contributed to high house prices. If you're more focused on potential value appreciation, bear in mind that while property can be a good investment, it isn't guaranteed to appreciate, especially in the short term.\\
    In conclusion, while no one can predict with certainty what will happen with the housing market over the next four years, it's crucial to make informed, careful decisions when considering buying property. It's advisable to consult with a professional to understand the potential risks and benefits based on your individual circumstances.}
\end{choruschat}

\paragraph{Conversation between U11 and \slackBot $N=5$.}
\begin{choruschat}
    \chatgptUser{This housing market is wild. Is it going to last the next 4 years?}
    \assistantA{Sure, I'd be happy to help with that. To get a better understanding of your situation, could you let me know if you're currently looking to buy or sell a property}
    \chatgptUser{I would like to buy a property}
    \assistantA{Understood. Could you let me know which area or city you're interested in buying property? This could help in providing a more accurate assessment.}
    \chatgptUser{[Location]}
    \assistantA{Thanks for sharing that. Now, are you looking to buy for investment purposes or for personal residence? This can influence the urgency and how market trends might impact your decision.}
    \chatgptUser{For personal residence}
    \assistantA{I see, thank you for sharing. Considering it's for personal residence, are you planning on living there long-term or is this a shorter term plan? }
    \chatgptUser{short term, maybe like 2 years}
    \assistantA{Thank you for the detail. Lastly, how flexible are you with the timing of your purchase? Would you be able to wait if it's recommended based on market predictions?}
    \chatgptUser{not really, I intend to  find a housing before fall ends}
    \assistantA{To answer your original question, predicting exact trends in the housing market can be notoriously difficult, as they can be influenced by a vast array of factors, including economic conditions, interest rates, and broader trends in population movement and growth.\\
    However, considering your specific situation - looking for a short-term personal residence in [Location] with an immediate purchase plan - here is a general perspective:\\
    Regardless of market trends over the next four years, your focus might be more on the current market conditions since your plan is short-term. [Location], like many places, has experienced a boost in the housing market recently, but this can change based on numerous factors.\\
    Given that you plan to purchase soon, it would be best to work with a local real estate agent who can provide specific advice tailored to your situation and the local [Location] market. They can help you navigate the current market conditions and find a property that suits your needs and budget.\\
    In conclusion, while the market's high activity might continue, it's crucial to focus on your immediate needs and current market conditions, especially given your short-term plan and immediate purchase intent.}    
\end{choruschat}

\paragraph{Conversation between U16 and \slackBot $N=7$.}
\begin{choruschat}
    \chatgptUser{This housing market is wild. Is it going to last the next 4 years?}
    \assistantA{Sure, I can help you with that. To start, can you tell me a bit about your current housing situation? Are you a homeowner, a renter, or looking to buy?}
    \chatgptUser{I am a renter right now.}
    \assistantA{That's helpful to know. Are you considering buying a home in the near future or are you comfortable with renting for now?}
    \chatgptUser{I'm still a student, so I am comfortable with renting.}
    \assistantA{Understood. May I ask why you are interested in the longevity of the current housing market? Are you considering investing in real estate or is it for other reasons?}
    \chatgptUser{I'm considering investing in real estate when I graduate.}
    \assistantA{Got it! That's an important future plan. Can you tell me when you'll be graduating and potentially ready to invest?}
    \chatgptUser{I'll  be graduate in two years, but considering invest in 5 years.}
    \assistantA{Thank you for sharing your timeline. Given that you're considering investing in a few years, what kind of real estate are you interested in? Residential properties, commercial, or something else?}
    \chatgptUser{ I'm interested in residential properties.}
    \assistantA{That's a popular choice for many first-time investors! In terms of location, are you considering investing in the area you're currently living in, or are you open to looking at properties in other locations as well? }
    \chatgptUser{I'm considering looking at properties in other locations as well, like in a big city.}
    \assistantA{That's a strategic approach, considering the potential for greater returns in a big city. Lastly, do you have any particular concerns about the housing market that you'd like me to address?}
    \chatgptUser{New York City may be great.}
    \assistantA{To answer your original question, predicting the exact trajectory of the housing market over the next four years is challenging due to the many variables at play. However, it's important to remember that the real estate market often moves in cycles, and while the current market may seem wild, it's normal for there to be periods of acceleration and slowdown.\\
    It's great that you're considering investing in residential properties, particularly in major cities like New York, where the market historically tends to rebound well even after downturns. However, keep in mind that the housing market can be influenced by many factors such as interest rates, economy, and population growth.\\
    As you'll only be ready to invest in about five years, keep an eye on the market trends, use this time to learn more about real estate investment, and prepare a strategy that aligns with your financial goals and risk tolerance. It’s encouraged to consult with a real estate professional or financial advisor when the time comes to make your investment.\\
    Remember, regardless of market conditions, a well-informed and carefully considered real estate investment has the potential to be a strong asset in the long run.}
\end{choruschat}

\section{MTurk interface for collecting worker responses}\label{sec:interface}
\begin{figure*}
    \centering
    \includegraphics[height=0.95\textheight]{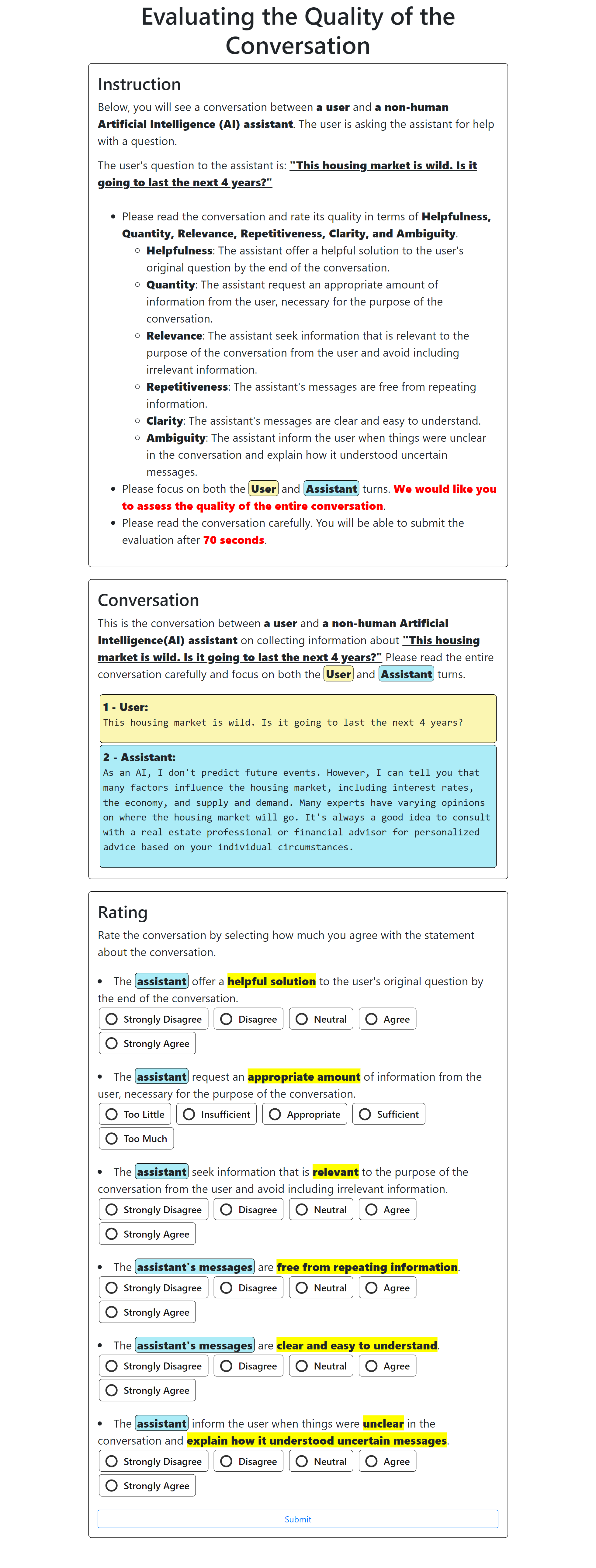}
    \caption{Interface for MTurk workers.}
    \Description{There are three separate sections. One on the top, one in the middle, and one at the bottom. The top section provides instructions for the MTurk workers accompanied with examples for each of the 8 inquiries. The middle section shows the target conversation. The bottom section are the 6 questionnaire questions we ask MTurk workers to provide their opinions.}
    \label{fig:annotation-interface}
\end{figure*}

\Cref{fig:annotation-interface} shows the MTurk interface for collecting online crowd workers' opinions through the six inquiries.
% \Cref{tab:category-rest} shows the remaining 17 open-coded categories.
% \Cref{tab:top_examples} shows example questions people would like to discuss in conversation (Conversation $\geq$ 4.22).
% interface_123

\end{document}